\documentstyle[aps,epsfig]{revtex}
\def\Id{{\rm 1\kern-.3em I}}
\def\gammagamma{{\gamma\gamma}}
\def\munu{{\mu\nu}}

\begin{document}

\title{\Large\bf\sc Compton Scattering off Relativistic Bound States: \\
	Two--Photon Vertices, Ward--Takahashi Identities, \\ Gauge
	Invariance and Low--Energy Limit}  

\author{ Matthias Koll\thanks{e-mail: {\tt koll@itkp.uni-bonn.de}} and Ralf Ricken
  \\ {\bigskip}  {\sl Institut f\"ur Theoretische Kernphysik \\
  Nu{\ss}allee 14--16, D--53115 Bonn \\ Germany} \\ {\bigskip} }


\maketitle

\begin{abstract}

In a general framework that has been labeled the ``gauging of
equations method'', we study the diagrams that contribute to Compton
scattering off a relativistic composite system. These contributions can be derived for
$N$--particle bound states described by the covariant Bethe--Salpeter
equation with a method
equivalent to minimal substitution in the one--particle case and yield the
correct contributions (including subtraction terms) in the order
${\cal O}(e^2)$. We give the Ward--Takahashi identities for the general
two--photon vertex as well as the corresponding constraints for the
two--photon irreducible interaction kernel and the Bethe--Salpeter
amplitude describing the bound state. From this we can 
show that gauge invariance holds for the full two--photon vertex. We
furthermore study in detail the low--energy limit of the Compton scattering
tensor in this approach (including a discussion of the pole terms) and
can prove that the full amplitude
yields the correct Born--Thomson limit as we shall explicitly
show for the spin--0 case. 

The calculations are completed by the investigation of certain
approximations that can be formulated for   
arbitrary $N$--particle bound states. We neglect for instance
contributions from $n$--photon irreducible interaction kernels and
show that in this case gauge invariance is only realized if either the interaction
kernel in the Bethe--Salpeter equation 
is independent of the total momentum and additionally is of local type,
or if the photon energies vanish; furthermore, we find the correct low--energy
limit in this approximation.
To clarify our approach, we also give the results in the order ${\cal O} (e)$;
as examples, we will quote some resulting lowest order expressions for a $q\bar q$
system explicitly.

\end{abstract}

\section{Introduction}

The first order result for a Green's function in an external field
obtained by a minimal substitution prescription and
the analoguous first order result for Bethe--Salpeter amplitudes have
been shown in refs. \cite{Blankleider1,Blankleider2,Blankleider3} by
Kvinikhidze and Blankleider; the authors refer    
to this procedure as ``gauging a hadronic system''. However, up to now
nothing is known about the important question how this general scheme can be
applied to Compton scattering off a hadronic system which is a
second order process. In fact, this is a difficult task since
electromagnetic gauge invariance is intimately related to the dynamics of the
strong interaction inside extended composite hadrons (see \cite{Ohta} and references therein). 
This important question has been addressed by Ito and Gross in ref. \cite{GrossIto} where
the authors have derived various diagrams that have to be included to
guarantee a gauge invariant description of the Compton scattering
process. These considerations --- based on a diagram analysis in a
general four--dimensional Bethe--Salpeter framework -- were explicitly
applied only to the deuteron system ({\it i.e.} with only one charged
constituent) as an example that keeps the equations as simple as possible.
In contrast to this approach, the procedure of gauging the relevant
bound state equations as applied here
has the advantage that it allows for a complete overview of the
possible contributions in a given order ${\cal O}(e^n)$ for any
$N$--particle  bound state; it thus provides a more
systematic view on the numerous processes contributing to Compton
scattering off a bound state. In this way,
it is therefore possible to study certain approximations (like
{\it e.g.} neglecting $n$--photon irreducible interaction kernels) on
the basis of the knowledge of the full expressions; this seems to be a
more firm way of introducing and handling these approximations than in the case
where one does not know what terms are precisely suppressed under
certain assumptions.

In this article, we aim --- on a general level --- at a complete
classification of the diagrams that contribute to the Compton
scattering process ${\cal A}\gamma\to{\cal A}\gamma$ for bound states
${\cal A}$ irrespective of their composition. Although our approach
can be considered as a quite technical study of this process, it seems
worthwhile to present it in this publication since it provides a more
transparent way of discussing this issue compared to the technically
involved explicit calculations in ref. \cite{GrossIto}; furthermore,
it allows for a simple check of the correct low--energy limits. We
thus demonstrate that the method of Blankleider and Kvinikhidze (see
\cite{Blankleider1,Blankleider2,Blankleider3}) turns 
out to be a powerful tool to derive all contributions not only in
first order of the electromagnetic interaction but also in the order
${\cal O}(e^2)$. Herein, the relevant
degrees of freedom are the constituent particles of the composite
system; our results for the full two--photon vertices turn out to be
independent of the interaction kernel used in the underlying
Bethe--Salpeter equation describing the bound state. In order to make clear
our notation and the scheme that we apply to the second order
amplitudes, we will always give the results in first order
({\it i.e.} ${\cal O}(e)$) for comparison. To simplify the notation, we will mostly
suppress the dependencies of relative momenta between the
constituents; however, we will give some explicit examples for $q\bar
q$ systems including the full dependency of the internal relative momentum.

After some introductory remarks on fundamental equations, definitions
and notations in section \ref{BetheSalpeter}, we 
will derive the second order expansions of Green's functions,
two--photon vertices and Bethe--Salpeter amplitudes in section
\ref{SecondOrder}. Ward--Takahashi identities and differential
expressions with respect to the bound state's total momentum are given in section
\ref{WardTakahashi} for vertices and
$n$--photon irreducible interaction kernels. The issue of gauge
invariance is addressed in section \ref{GaugeInvariance} before we
study the low--energy limit of the Compton scattering tensor in
section \ref{LowEnergyLimit}. In section \ref{Summary}, we summarize
our results.

\section{Green's Functions, T--Matrix and Bethe--Salpeter Equations} 
                                        \label{BetheSalpeter}

The fundamental Green's function equation for a given number of $n$ fermions
and $\bar n$ anti-fermions --- depicted in
fig. (\ref{fig:GreensFunctionEquation}) in a graphical form --- is
defined as follows:  

\begin{eqnarray}
\label{GreensFunctionEquation}
G &=& G_0 - i G_0 K G  \\
  &=& G_0 - i G K G_0  \nonumber \quad .
\end{eqnarray}

\protect\begin{figure}[h]
  \protect\begin{center}
    \leavevmode
       \protect\input{GreensEquation.pstex_t}
\newline
       \protect\caption{The equation for the Green's function $G$
       according to eq. (\protect\ref{GreensFunctionEquation}). The
       vertical dots (here and in the following figures) denote the
       possible occurence of additional fermion lines depending on the
       number of constituents.}
       \protect\label{fig:GreensFunctionEquation}
   \end{center}
\end{figure}

Throughout this article, all $n+\bar n - 1$ four--dimensional integrations
over internal relative momenta are implicitly understood;
furthermore, we omit all indices for internal quantum numbers such as
flavour, colour and spin. To 
clarify this point, we write down the fundamental Green's equation,
{\it i.e.} the first line in eq. (\ref{GreensFunctionEquation}), for
$G\left(P;\{k_j^\prime\}; \{k_j\}\right)=G\left(\{p_i^\prime\};
\{p_i\}\right)$ with $i=1,2,\dots, N$ and $j=1,2,\dots, N-1$ explicitly as

\begin{eqnarray}
\label{GreensFunctionEquationExplicit}
\lefteqn{\left[G\left(P;\{k_j^\prime\}; \{k_j\}\right)\right]_{\{\alpha_i\}
, \{\beta_i\}} =
\left[G_0\left(P;\{k_j^\prime\}; \{k_j\}\right)\right]_{\{\alpha_i\}
, \{\beta_i\}} }\\ 
&-& i \int\frac{d^4\:\{q_j^\prime\}}{\left(2\pi\right)^{4(N-1)}}
\int\frac{d^4\:\{q_j\}}{\left(2\pi\right)^{4(N-1)}}\:    
\left[G_0\left(P;\{k_j^\prime\};
\{q_j^\prime\}\right)\right]_{\{\alpha_i\},
 \{\gamma_i\}}\left[K\left(P;\{q_j^\prime\};
\{q_j\}\right)\right]_{\{\gamma_i\},
 \{\kappa_i\}}\left[G\left(P;\{q_j\}; \{k_j\}\right)\right]_{\{\kappa_i\}
, \{\beta_i\}} \nonumber
\end{eqnarray}

where we have separated the total momentum $P=p_1+\dots+p_N=p_1'+\dots+p'_N$
($N=n+\bar n$) and we integrate over all relative variables $\{k_j\}=k_1,
k_2,\dots, k_{N-1}$; see also appendix \ref{app:Coordinates}. Here,
the index set $\{\alpha _i\}=\alpha _1, \ldots, \alpha _N$ represents the
multi--indices for each constituent particle in flavour, colour and spin space. Since
$G_0$ describes free propagation without any interaction between the
constituents, it is defined as 

\begin{eqnarray}
G_0\left(P;\{k_j^\prime\}; \{k_j\}\right) := G_0\left(P;
\{k_j\}\right) \cdot \left(2\pi\right)^4\delta 
^4\left(k_1^\prime - k_1\right)\dots\left(2\pi\right)^4\delta
^4\left(k_{N-1}^\prime - k_{N-1}\right)\nonumber
\end{eqnarray}

in this explicit formula. Here and in what follows in this section,
all quantities are functions of the total momentum $P$; furthermore,
the dependence on the relative momenta $k_j$ of the constituents
(aside from some examples) and all indices will be suppressed in this paper. 

The kernel $K$ contains all irreducible interaction terms
between fermions and anti--fermions, see
fig. (\ref{fig:IrrKernel}). In this publication, we will not make any
assumption about this kernel; however, we will show in section
\ref{KernelWTI} that it has to obey certain -- general --- Ward--Takahashi identities
depending on the approximations applied in order to preserve gauge
invariance for electromagnetic processes. 

\protect\begin{figure}[h]
  \protect\begin{center}
    \leavevmode
       \protect\input{Kernel.pstex_t}
\newline
       \protect\caption{Irreducible diagrams that contribute to the
       interaction kernel $G_0 KG_0$ with $N=4$ constituents as an example.}
       \protect\label{fig:IrrKernel}
   \end{center}
\end{figure}

Let us stress that this kernel $K$ includes not only $N$--particle
irreducible diagrams but also diagrams in which some of the
constituents are unaffected by the interactions between the other
constituents; an example is given in the second diagram in
fig. (\ref{fig:IrrKernel}). In this sense, the kernel $K$ in the Green's
function equation is the sum of all possible $i$--particle irreducible kernels
$K^{(i)}$ ($i=2,\dots,N$) (see {\it e.g.} \cite{Loering} for a precise
formulation of $K=\sum _i K^{(i)}$ in the case of $qqq$ baryons). 

The iteration of an infinite sum of irreducible diagrams in
eq. (\ref{GreensFunctionEquation}) can also be written as $G=G_0-iT$ 
by introducing the $T$ matrix as $T := G_0KG$ which then satisfies the
following equations: 

\begin{eqnarray}
\label{TMatrixEquation}
T &=& G_0KG_0 - i G_0 K T \\
  &=& G_0KG_0 - i T K G_0 \quad .\nonumber
\end{eqnarray}

\protect\begin{figure}[h]
  \protect\begin{center}
    \leavevmode
       \protect\input{TMatrixEquation.pstex_t}
\newline
       \protect\caption{The equation for the $T$ matrix
       in eq. (\protect\ref{TMatrixEquation}).}
       \protect\label{fig:TMatrixEquation}
   \end{center}
\end{figure}

From the equation for the Green's function, we can deduce the
Bethe--Salpeter equation for the Bethe--Salpeter amplitude $\chi $
describing on--shell bound states and its adjoint $\bar \chi $ (see
 \cite{SalpeterBethe,Lurie}):

\begin{eqnarray}
\label{BetheSalpeterEquation1}
\chi  = - i G_0 K \chi  \qquad \mbox{and}\qquad\bar\chi  = - i \bar\chi  K G_0 \quad .
\end{eqnarray}

Here, $\chi$ is a short--hand notation for the
Bethe--Salpeter amplitude $\chi _{\alpha
_1,\ldots,\alpha _{n+\bar n}}  (P;\{k_j\})$ that in coordinate space
is defined explicitly as

\begin{eqnarray}
\label{DefBSA}
\chi ^{P}_{\alpha _1,\ldots,\alpha _{n+\bar n}}
(x_1,\ldots, x_{n+\bar n}) &:=&  \left\langle\: 0 \: \left |\:
 \textsf{T} \:\psi_{\alpha_1} (x_1)\ldots\psi_{\alpha_n}
 (x_n)\bar\psi_{\alpha _{n+1}} (x_{n+1})\ldots\bar\psi_{\beta
 _{n+\bar n}} (x_{n+\bar n})\right |\:P \:\right\rangle  \\ 
&=& e^{-iPX} \int\frac{d^4\:\{k_j\}}{\left(2\pi\right)^{4(N-1)}}\:
 e^{-ik_1 r_1}\ldots e^{-ik_{N-1} r_{N-1}} \chi _{\alpha
_1,\ldots,\alpha _{n+\bar n}}  (P;\{k_j\}) \nonumber
\end{eqnarray}

for a bound state of $n$ fermions and $\bar n$
anti--fermions ($N=n+\bar n$). In this definition,
$\textsf{T}$ denotes the time--ordered product; see also appendix
\ref{app:Coordinates} for the definition of the coordinates
$X,r_1,\ldots,r_{N-1}$.   

With the formal solution of eq. (\ref{GreensFunctionEquation}), {\it i.e.}
$G = (G^{-1}_0+iK)^{-1}$, the Bethe--Salpeter equations in
eq. (\ref{BetheSalpeterEquation1}) can be formulated as follows:

\begin{eqnarray}
\label{BetheSalpeterEquation2}
G^{-1}\chi  = 0 \qquad \mbox{and}\qquad\bar\chi G^{-1} =0\quad .
\end{eqnarray}

Note that these homogeneous Bethe--Salpeter equations are only valid (and thus the
Bethe--Salpeter amplitudes are only defined) for on--shell bound 
states of momentum $P$ and mass $M$ with the constraint $P^2=M^2$.

\protect\begin{figure}[h]
  \protect\begin{center}
    \leavevmode
       \protect\input{BetheSalpeterEquation.pstex_t}
\newline
       \protect\caption{The Bethe--Salpeter equation for a bound
    states of $N$ constituents in a graphical
    notation, see eq. (\ref{BetheSalpeterEquation1}).}
       \protect\label{fig:BSE}
   \end{center}
\end{figure}

\section{Second Order Expansion in the Electromagnetic Field}
                                        \label{SecondOrder} 

In the following we extend the fundamental Green's equation and the
Bethe--Salpeter equation by expanding operators $\textsf{O}=G_0, G, K,
\chi$ in the presence of an external electromagnetic field as

\begin{eqnarray}
\label{Expansion}
 \textsf{O}^A &=& \textsf{O} - ie\textsf{O}^\gamma - e^2
\textsf{O}^{\gamma\gamma} + {\cal O}(e^3) \quad .
\end{eqnarray}

In this way, we find the expressions for the second order Green's
function $G^\gammagamma$ and the second order Bethe--Salpeter
amplitude $\chi ^\gammagamma$.

\subsection{Green's Functions}
                                         \label{GreensFunctions} 

The Green's function in eq. (\ref{GreensFunctionEquation}) can be
transformed into a Green's function in an external field via 
minimal substitution:

\begin{eqnarray}
\label{GreensFunctionEquationWithA}
G^A &=& G^A_0 - i G^A_0 K^A G^A 
\end{eqnarray}

where eq. (\ref{Expansion}) for the operators $\textsf{O}=G_0, G, K$ is
applied. In this way, each fermion line in the 
operators $\textsf{O}$ will be attached by a number of photons in the
order ${\cal O}(e^n)$ similar to the well--known construction of the
fermion propagator $S^A(p,p')$ in an electromagnetic field, see
fig. (\ref{fig:FermionPropagator}).

\protect\begin{figure}[h]
  \protect\begin{center}
    \leavevmode
       \protect\input{FermionPropagator.pstex_t}
\newline
       \protect\caption{The fermion propagator in an external field.}
       \protect\label{fig:FermionPropagator}
   \end{center}
\end{figure}

\subsubsection{First Order Expansion}

With the expansion in eq. (\ref{GreensFunctionEquationWithA}) and by
using eq. (\ref{GreensFunctionEquation}), 
we can write down an equation for the Green's function in first order
of the electromagnetic coupling:

\begin{eqnarray}
\label{GreensFunction1Order1}
G^\gamma &=& G^\gamma_0 - i G^\gamma_0 K G  - i G_0 K^\gamma G  - i G_0 K G^\gamma\nonumber \\
 &=& \underbrace{\big(\Id+iG_0K\big)^{-1}}_{=GG_0^{-1}}\big(
 G_0^\gamma\underbrace{(\Id-iKG)}_{= G_0^{-1}G}  - i G_0 K^\gamma
 G\big) \nonumber \\
 &=& G\left(G_0^{-1}G^\gamma_0 G_0^{-1} - i K^\gamma\right)G \nonumber\quad .
\end{eqnarray}

Defining the amputated full one--photon vertex
$\Gamma^\gamma:=G^{-1}G^\gamma G^{-1}$ and the amputated free photon
vertex $\Gamma_0^\gamma:=G_0^{-1}G_0^\gamma G_0^{-1}$, we can write

\begin{eqnarray}
\label{GreensFunction2Order1}
G^\gamma = G\Gamma^\gamma G \qquad \mbox{with}\quad  \Gamma^\gamma =
\Gamma_0^\gamma - i K^\gamma
\end{eqnarray}

for the first order Green's function in an external field; this
equation is depicted diagrammatically in fig. (\ref{fig:GreensFunction1Order}).

\protect\begin{figure}[h]
  \protect\begin{center}
    \leavevmode
       \protect\input{Gamma1.pstex_t}
\newline
       \protect\caption{The one--photon vertex $G_0\Gamma^\gamma G_0$
       according to eq. (\ref{GreensFunction2Order1}) in a graphical notation.}
       \protect\label{fig:GreensFunction1Order}
   \end{center}
\end{figure}

This full one--photon vertex can not solely be studied in models where
the Bethe--Salpeter equation is used in its instantaneous
approximation; it is also relevant in models
including retardation effects (see \cite{WallacePhillips}). The
concept of an ``interaction current'' $K^\gamma$ has been
introduced in ref. \cite{GrossRiska}; its effects have been studied
quantitatively in \cite{ItoBuckGross2} where the 
electromagnetic pion form factor is calculated with the full vertex
$\Gamma^\gamma$ in a four--dimensional
separable ansatz for $K$ based on `t Hooft's instanton--induced
interaction (see also \cite{ItoBuckGross1}).

To make clear the physical meaning of $K^\gamma$, we give some
diagrams contributing to this kernel explicitly in
fig. (\ref{fig:KernelGamma}) as an example.   
Note that --- in analogy to the kernel without electromagnetic field
in fig. (\ref{fig:IrrKernel}) --- all $i$--particle irreducible
diagrams ($i=2,\dots,N$) contribute to $K^\gamma$, {\it i.e.} also
processes where some of the constituents can be regarded as spectators.

\protect\begin{figure}[h]
  \protect\begin{center}
    \leavevmode
       \protect\input{KernelGamma.pstex_t}
\newline
       \protect\caption{The one--photon irreducible interaction kernel
       $G_0 K^\gamma G_0$ with $N=4$ charged constituents as an example. If
       the interaction kernel $K$ includes charged exchange particles
       then additional diagrams contribute to $K^\gamma$ in which the
       photon also couples to these charged exchange particles.}
       \protect\label{fig:KernelGamma}
   \end{center}
\end{figure}

For bound states composed of more than two constituents ({\it e.g.} a
nucleon described as a $qqq$ system), the method of gauging the
Green's function equation provides a solution to the notorious problem
of overcounting diagrams in few--body systems. A detailed discussion
of the role of subtraction terms emerging in this framework can be
found in refs. \cite{Blankleider2,Blankleider3} where the problem of
ambiguous cuts in the related diagrams (as well as its cure) is
intensively studied.

We remark moreover that our notion of a ``lowest order vertex'' with
respect to $\Gamma^\gamma _0$ (and as well to $\Gamma^\gammagamma _0$
introduced in the next paragraph) is somewhat sloppy since --- apart
from coupling constants that might be absorbed in the interaction
kernels $K$, $K^\gamma$ and $K^\gammagamma$ --- we have no
ordering or power counting scheme that is reflected in this term. In
principal, all parts of $\Gamma^\gamma$ (and $\Gamma^\gammagamma$) can
contribute in the same magnitude to the full $n$--photon irreducible
vertices. However, these lowest order vertices are minimal in the
sense that they should be included in any framework that considers
bound states like the pion or the nucleon as composite systems since
they describe the simplest photon couplings to the
constituents. Therefore, we will use this phrase now and then in the
following discussion although it is not strictly exact.

\subsubsection{Second Order Expansion}

For the description of Compton scattering off a bound state described
by the Bethe--Salpeter equation, we need the second order expansion of
the Green's function. Let us therefore again start with
eq. (\ref{GreensFunctionEquationWithA}) and evaluate it in the order
${\cal O}(e^2)$:

\begin{eqnarray}
\label{GreensFunction1Order2}
G^\gammagamma &=& G^\gammagamma_0 - i G^\gammagamma_0 K G  - i G_0
 K^\gammagamma G  - i G_0 K G^\gammagamma - i G^\gamma_0 K^\gamma G  - i G^\gamma_0
 K G^\gamma  - i G_0 K^\gamma G^\gamma\nonumber \\
 &=& \underbrace{\big(\Id+iG_0K\big)^{-1}}_{=GG_0^{-1}}\big(
 G_0^\gammagamma\underbrace{(\Id-iKG)}_{= G_0^{-1}G}  - i G_0 K^\gammagamma
 G- i G^\gamma_0 K^\gamma G  - i G^\gamma_0
 K G^\gamma  - i G_0 K^\gamma G^\gamma\big) \nonumber \\
 &=& G\big(G_0^{-1}G^\gammagamma_0 G_0^{-1} - i K^\gammagamma \underbrace{- i
 G_0^{-1}G_0^\gamma K^\gamma - i G_0^{-1}G_0^\gamma K G^\gamma G^{-1}-
 i K^\gamma G^\gamma G^{-1}}_{=:A^\gammagamma}\big)G \nonumber\quad .
\end{eqnarray}

The last three terms can be simplified with the expression for the
full one--photon vertex in eq. (\ref{GreensFunction2Order1}):  

\begin{eqnarray}
A^\gammagamma &=&- i G_0^{-1}G_0\Gamma_0^\gamma  G_0 K^\gamma - i
 G_0^{-1}G_0\Gamma_0^\gamma \underbrace{G_0 K G }_{i(G-G_0)}\Gamma^\gamma GG^{-1} -
 i K^\gamma G \Gamma^\gamma GG^{-1} \nonumber \\
&=&- i \Gamma_0^\gamma  G_0 K^\gamma + \Gamma_0^\gamma G \Gamma^\gamma
 - \Gamma_0^\gamma G_0 (\Gamma_0^\gamma - i K^\gamma) - i K^\gamma G
 \Gamma^\gamma  \nonumber\\ 
&=& \Gamma^\gamma G \Gamma^\gamma - \Gamma_0^\gamma G_0 \Gamma_0^\gamma \quad . \nonumber
\end{eqnarray}

We define --- in analogy to the first order expansion --- the
amputated full two--photon vertex 
$\Gamma^\gammagamma:=G^{-1}G^\gammagamma G^{-1}$ and the amputated free
two--photon vertex $\Gamma_0^\gammagamma:=G_0^{-1}G_0^\gammagamma
G_0^{-1}$; the second order Green's function then reads 

\begin{eqnarray}
\label{GreensFunction2Order2}
G^\gammagamma = G\Gamma^\gammagamma G \qquad \mbox{with}\quad  \Gamma^\gammagamma =
\Gamma_0^\gammagamma - i K^\gammagamma + \Gamma^\gamma G \Gamma^\gamma
- \Gamma_0^\gamma G_0 \Gamma_0^\gamma 
\end{eqnarray}

and is shown in terms of diagrams in
fig. (\ref{fig:GreensFunction2Order}). This is the key result of this
section since it provides --- on the basis of rather simple
considerations --- the answer to the question which processes
contribute to Compton scattering off a bound state. In this sense,
eq. (\ref{GreensFunction2Order2}) can be regarded as a generalization
of the results of ref. \cite{GrossIto}; furthermore and in addition to
it, it gives a more systematic overview and ordering of the various
terms and by this means thus reveals the existence of a subtraction
term besides the pole contributions in $\Gamma^\gamma G \Gamma^\gamma$.
Note that this subtraction term $\Gamma_0^\gamma G_0 \Gamma_0^\gamma$
naturally emerges in this framework; it prevents overcounting of the
lowest--order parts in $\Gamma^\gamma G\Gamma^\gamma$ and turns out to
be crucial for a proper description of Compton scattering off bound
states. The two--photon irreducible interaction kernel $K^\gammagamma$
in eq. (\ref{GreensFunction2Order2}) can be understood analogously to 
fig. (\ref{fig:KernelGamma}). 

\protect\begin{figure}[h]
  \protect\begin{center}
    \leavevmode
       \protect\input{Gamma2.pstex_t}
\newline
       \protect\caption{The two--photon vertex $G_0\Gamma^\gammagamma G_0$
       according to eq. (\ref{GreensFunction2Order2}) in a graphical notation.}
       \protect\label{fig:GreensFunction2Order}
   \end{center}
\end{figure}

As an approximation of this full two--photon vertex, we will introduce
$\hat G^\gammagamma := G \hat\Gamma^\gammagamma G$ where all
$n$--photon irreducible interaction kernels, {\it i.e.} $K^\gamma$ and
$K^\gammagamma$, are neglected. Then, by virtue of $G=G_0-iT$, the subtraction term can be
evaluated explicitly --- see fig. (\ref{fig:HatGreensFunction2Order})
--- finally yielding

\begin{eqnarray}
\label{HatGreensFunction2Order2}
\hat G^\gammagamma = G\hat\Gamma^\gammagamma G \qquad \mbox{with}\quad
\hat\Gamma^\gammagamma = \Gamma_0^\gammagamma - i \Gamma_0^\gamma T
\Gamma_0^\gamma \quad .
\end{eqnarray}

\protect\begin{figure}[h]
  \protect\begin{center}
    \leavevmode
       \protect\input{HatGamma2.pstex_t}
\newline
       \protect\caption{The approximated two--photon vertex $G_0\hat\Gamma^\gammagamma G_0$
       without inclusion of $n$--photon irreducible interaction
       kernels according to eq. (\ref{HatGreensFunction2Order2}) in a graphical notation.}
       \protect\label{fig:HatGreensFunction2Order}
   \end{center}
\end{figure}

\subsubsection{Explicit Formulae for Compton Scattering}

These results in first and second order of the coupling $e$ were formulated for a
hadron in an amorphous electromagnetic field without stating precisely
the photon momenta and coupling indices in Dirac space. If we consider
Compton scattering kinematics ({\it i.e.} $P+q_1=P'+q_2$), we now have to
indicate not only the correct momentum transfers but also to specify
coupling indices $\mu, \nu,\ldots$ of the
photon lines and make sure that in each order all possible diagrams including
permutations are taken into account. For the second order fermion
propagator, this yields two diagrams (their sum being explicitly crossing symmetric)
as shown in fig. (\ref{fig:FermionPropagator2ndOrder}).

\protect\begin{figure}[h]
  \protect\begin{center}
    \leavevmode
       \protect\input{FermionPropagator2.pstex_t}
\newline
       \protect\caption{The fermion propagator in an external field in
       second order with Compton scattering kinematics $P'=P+q_1-q_2$
       for two photons $\gamma _1=\gamma
       _1(q_1,\mu)$ and $\gamma _2=\gamma _2(-q_2,\nu)$.}
       \protect\label{fig:FermionPropagator2ndOrder}
   \end{center}
\end{figure}

For the first order Green's function in
eq. (\ref{GreensFunction2Order1}), all terms can be identified
uniquely and therefore no correction is needed:

\begin{eqnarray}
\label{GreensFunction2Order1mu}
G^\mu (P',P) &=& G(P')\Gamma^\mu (P',P)G(P) \\ \mbox{with}\quad  \Gamma^\mu (P',P)&=&
\Gamma_0^\mu(P',P) - i K^\mu(P',P) \nonumber\quad .
\end{eqnarray}

Here we have introduced the momentum transfer $P'=P+q$ for the
photon $\gamma=\gamma (q,\mu)$ involved in this process.

Considering the second order Green's function in
eq. (\ref{GreensFunction2Order2}), we find that additional terms
appear due to crossing symmetry:

\begin{eqnarray}
\label{GreensFunction2Order2munu}
G^\munu (P',P) &=& G(P')\Gamma^\munu(P',P) G (P) \\
\mbox{with}\quad  \Gamma^\munu (P',P) &=&
\Gamma_0^\munu(P',P) - i K^\munu(P',P) \nonumber\\
&+&\Gamma^\mu (P',P'-q_1) G
(P\:-q_2)\Gamma^\nu (P-q_2,P) - \Gamma_0^\mu (P',P'-q_1) G_0
(P\:-q_2)\Gamma_0^\nu (P-q_2,P) \nonumber \\
&+& \Gamma^\nu(P',P'+q_2) G
(P'+q_2)\Gamma^\mu(P+q_1,P) - \Gamma_0^\nu(P',P'+q_2) G_0
(P'+q_2)\Gamma_0^\mu(P+q_1,P)\nonumber \quad .
\end{eqnarray}

For Compton scattering with the incoming photon $\gamma _1=\gamma       
_1(q_1,\mu)$ and the outgoing photon $\gamma _2=\gamma _2(-q_2,\nu)$,
energy--momentum conservation requires $P+q_1=P'+q_2$ where $P$ ($P'$)
denotes the incoming (outgoing) total four--momentum of the bound
state. After multiplying out all contributions from $\Gamma^\gamma G
\Gamma^\gamma$, all terms in
eq. (\ref{GreensFunction2Order2munu}) can be identified with diagrams
shown in \cite{GrossIto} where Compton scattering off a deuteron
system ({\it i.e.} with only one charged constituent) has been studied.

It will be helpful for some following remarks to split up the Compton
vertex in eq. (\ref{GreensFunction2Order2munu}) into parts that
include the pole terms induced by the full Green's function $G$ and
the rest that is non--singular (see {\it e.g.}
\cite{SchererFearing,GellMannGoldberger}): 

\begin{eqnarray}
\label{VertexSplit}
\Gamma^\munu (P',P) &=& \Gamma^\munu_{\mbox{\footnotesize\sc Pole}} (P',P) +
\Gamma^\munu_{\mbox{\footnotesize\sc NoPole}} (P',P)\\ 
\mbox{with}\quad  \Gamma^\munu_{\mbox{\footnotesize\sc Pole}} (P',P) &=&
\Gamma^\mu (P',P'-q_1) G 
(P\:-q_2)\Gamma^\nu (P-q_2,P) \nonumber \\
&+& \Gamma^\nu(P',P'+q_2) G
(P'+q_2)\Gamma^\mu(P+q_1,P) \nonumber\quad .
\end{eqnarray}

Let us note that from the equations above no constraint arises
neither for the one--photon irreducible interaction kernel
$K^\mu(P',P)$ nor for the two--photon irreducible interaction kernel 
$K^\munu(P',P)$. However, they have to obey certain
Ward--Takahashi identities in order to satisfy the constraints from
gauge invariance as we will see in section \ref{KernelWTI}. Concerning
the interpretation of the $n$--photon irreducible 
interaction kernels, it is worthwile to note that one can generally describe
them in a different way than demonstrated above. Let us therefore
first define 

\begin{eqnarray}
G_{(\epsilon)}  &:=& G_0 - i\epsilon\: G_0 K G 
\end{eqnarray}

as a modified Green's function that reduces to $G_{(0)}= G_0$
describing free propagation for $\epsilon =0$. With this definition,
we introduce

\begin{eqnarray}
\Gamma_{(\epsilon)}       &:=& G_0^{-1} + i\epsilon\:  K \\
\Gamma^\mu_{(\epsilon)}   &:=& \Gamma_0^\mu - i\epsilon\:  K^\mu\nonumber\\
\Gamma^\munu_{(\epsilon)} &:=& \Gamma_0^\munu - i\epsilon\: K^\munu
+\Gamma^\mu_{(\epsilon)}G_{(\epsilon)}\Gamma^\nu_{(\epsilon)} +
\Gamma^\nu_{(\epsilon)}G_{(\epsilon)}\Gamma^\mu_{(\epsilon)}
-\Gamma^\mu_{0}G_{0}\Gamma^\nu_{0} - \Gamma^\nu_{0}G_{0}\Gamma^\mu_{0}\nonumber
\end{eqnarray}

and then finally 

\begin{eqnarray}
\label{CalGDef}
{\cal G}_{(\epsilon)}       &:=& \Gamma_{(\epsilon)}\\
{\cal G}^\mu_{(\epsilon)}   &:=& \Gamma_{(\epsilon)}^\mu\nonumber\\
{\cal G}^\munu_{(\epsilon)} &:=& \Gamma_{(\epsilon)}^\munu -
\Gamma_{(\epsilon)}^\mu G_{(\epsilon)}\Gamma_{(\epsilon)}^\mu -
\Gamma_{(\epsilon)}^\nu G_{(\epsilon)}\Gamma_{(\epsilon)}^\nu \nonumber
\end{eqnarray}

which are quantities that are free from singularities. For $\epsilon =
0$, they obviously include no contributions from interaction kernels
$K$, $K^\mu$ and $K^\munu$ while for $\epsilon = 1$ we recover ${\cal
G}_{(1)}=G^{-1}$, ${\cal G}^\mu_{(1)}=\Gamma^\mu$ and ${\cal
G}_{(1)}^\munu=\Gamma^\munu_{\mbox{\footnotesize\sc
NoPole}}=\Gamma^\munu-\Gamma^\munu_{\mbox{\footnotesize\sc Pole}}$. With
these definitions, one can easily verify that the interaction kernels
can simply be written as     
 
\begin{eqnarray}
\label{KWithCalG}
+iK &=& {\cal G}_{(1)} - {\cal G}_{(0)} \\
\label{KmuWithCalG}
-iK^\mu &=& {\cal G}^\mu_{(1)} - {\cal G}^\mu_{(0)} \\
\label{KmunuWithCalG}
-iK^\munu &=& {\cal G}^\munu_{(1)} - {\cal G}^\munu_{(0)} 
\end{eqnarray}

We will come back to the definitions above in section \ref{GreensWTI}
since they obey Ward--Takahashi identities in a unified way
due to their lack of poles; moreover, a derivation of Ward--Takahashi
identities for the $n$--photon interaction kernels is straightforward
with the help of eqs. (\ref{KWithCalG})--(\ref{KmunuWithCalG}).

\subsection{Bethe--Salpeter Amplitudes}
                                    \label{BetheSalpeterAmplitudes} 

With the systematic expansion in eq. (\ref{Expansion}), it is also
possible to derive an ``extended'' Bethe--Salpeter equation 

\begin{eqnarray}
\label{BSExpansion}
\chi ^A = - i G^A_0 K^A \chi ^A 
\end{eqnarray}

for a bound state in an external field; here, we expand $\textsf{O}=G_0, K, \chi$. For the
first order result, one can easily show that

\begin{eqnarray}
\label{BSA1stOrder}
\chi ^\gamma = G \Gamma^\gamma \chi 
\end{eqnarray}

(see also \cite{Blankleider1}). With a little algebra, we find the
second order result (see appendix \ref{app:BSA2}):

\begin{eqnarray}
\label{BSA2ndOrder}
\chi ^\gammagamma = G \Gamma^\gammagamma \chi \quad .
\end{eqnarray}

Similar relations hold for the adjoint amplitudes if
we start with the ``extended'' adjoint Bethe--Salpeter equation $\bar\chi
^A = - i \bar\chi ^A K^A G^A_0$: 

\begin{eqnarray}
\label{barBSA1stOrder}
\bar\chi ^\gamma &=&\bar\chi\Gamma^\gamma  G  \quad , \\
\label{barBSA2ndOrder}
\bar\chi ^\gammagamma &=& \bar\chi\Gamma^\gammagamma G  \quad .
\end{eqnarray}

For the sake of completeness, we finally want to give these results 
for distinguishable photons:

\begin{eqnarray}
\label{BSAmu}
\chi ^\mu (P',P)&=& G (P')\:\Gamma^\mu (P',P)\:\chi(P) \quad , \\
\label{BSAmunu}
\chi ^\munu (P',P)&=& G (P')\Gamma^\munu (P',P)\chi(P)  
\end{eqnarray}

in first and second order, respectively, and analogously for the
adjoint amplitudes $\bar\chi ^\mu (P',P)$ and $\bar\chi ^\munu (P',P)$. 
In this sense, the expanded (adjoint) Bethe--Salpeter amplitudes $\chi^\mu$ and
$\chi^\munu$ ($\bar\chi ^\mu$ and $\bar\chi ^\munu$) are no ``new''
quantities obeying separately relations 
like $n$-th order (adjoint) Bethe--Salpeter equations but they can be formulated in terms of
the full Green's function, the full $n$--photon vertex and a
``genuine'' or ``free'' (adjoint) Bethe--Salpeter amplitude without
any coupling to the electromagnetic field.

Let us stress that eqs. (\ref{BSAmu}) and (\ref{BSAmu}) represent well
defined relations since the Green's function $G$ remains finite for
$q=P'-P\not= 0$; we will come back to this point later.

\section{Ward--Takahashi Identities}
                                        \label{WardTakahashi} 

Ward--Takahashi identities (see \cite{Ward,Takahashi}) relate
amplitudes of electromagnetic processes of order ${\cal O}(e^{n+1})$
with those of order ${\cal O}(e^n)$. In the following, we will derive
constraints for the one--photon and two--photon irreducible interaction
kernels from these relations as well as identities for the
Bethe--Salpeter amplitudes in first and second order of the
electromagnetic coupling. 

Since we aim at a general description of bound states composed of an
unspecified number of $N$ constituents, we will again in general suppress all
relative momenta and indicate only the dependence of the total
four--momenta $P$ and $P'$; explicit examples of Ward--Takahashi
identities for a $q\bar q$ system are given in section \ref{ExamplesForMesons}.

\subsection{Identities for the Amputated Green's Functions}
                                    \label{GreensWTI} 

In \cite{PeskinSchroeder}, a general form of the celebrated
Ward--Takahashi identity is given (see also \cite{Kazes} and the
seminal publications \cite{Ward,Takahashi}):

\begin{eqnarray}
\label{PeskinWTI}
q_\mu M^\mu (P',P) = e \Big( M (P'-q,P) - 
M(P',P+q)\Big)\quad .
\end{eqnarray}

Here, $M^\mu$ denotes the amplitude for the process with $n+1$
photons and $M$ is the amplitude for the process with $n$
photons; the indices for these $n$ other photons as well as the
dependency of $M^\mu$ and $M$ of the internal relative momenta are suppressed in
this notation. In fig. (\ref{fig:PeskinWTI}), this formula is depicted
diagrammatically.

\protect\begin{figure}[h]
  \protect\begin{center}
    \leavevmode
       \protect\input{WardTakahashi.pstex_t}
\newline
       \protect\caption{The most general form of the Ward--Takahashi
       identity for a photon $\gamma=\gamma(q,\mu)$ according to
       eq. (\ref{PeskinWTI}); see also \protect\cite{PeskinSchroeder}.} 
       \protect\label{fig:PeskinWTI}
   \end{center}
\end{figure}

The amplitudes for processes with zero, one and two photons involved
are given by

\begin{eqnarray}
\label{AmplitudesM}
\begin{array}{rclccc}
M(P',P) &:=&        & G(P') & \Gamma (P',P) & G(P) \\
M^\mu(P',P) &:=& (-ie)      & G(P') & \Gamma^\mu (P',P) & G(P) \\
M^\munu(P',P) &:=&   (-ie)^2     & G(P') & \Gamma ^\munu(P',P) & G(P) 
\end{array}\quad ,
\end{eqnarray}

respectively. The vertices (or amputated Green's functions) $\Gamma^\mu$ and
$\Gamma^\munu$ are given in eqs. (\ref{GreensFunction2Order1mu}) and
(\ref{GreensFunction2Order2munu}). Furthermore, we have defined
$\Gamma(P',P):=G^{-1}(P)\cdot(2\pi)^4\delta ^4(P'-P)$; for the charge
factors in the equations above, see section \ref{ChargeFactors}.

For the electromagnetic current in the order ${\cal O}(e)$ with $P'=P+q$, the Ward--Takahashi
identity reads explicitly

\begin{eqnarray}
\label{WTImu}
q_\mu M^\mu (P+q,P) &=& e \Big( M(P,P) - 
M(P+q,P+q)\Big) \nonumber\\
\Longleftrightarrow \qquad -iq_\mu\Gamma^\mu(P+q,P) &=& G^{-1}(P+q) -
G^{-1}(P) \quad .
\end{eqnarray}

In the case of Compton scattering, four--momentum conservation
requires $P+q_1 = P'+q_2$; the Ward--Takahashi identity for this
process  in order ${\cal O}(e^2)$ is therefore 

\begin{eqnarray}
\label{WTImunu}
q_{1\mu} M^\munu (P+q_1-q_2,P) &=& e \Big( M^\nu(P-q_2,P) - 
M^\nu(P+q_1-q_2,P+q_1)\Big) \nonumber\\
\Longleftrightarrow \qquad -iq_{1\mu}\Gamma^\munu(P+q_1-q_2,P) &=&
G^{-1}(P+q_1-q_2)G(P-q_2)\Gamma^\nu(P-q_2,P) \nonumber\\
&-& \Gamma^\nu(P+q_1-q_2,P+q_1)G(P+q_1)G^{-1}(P) \quad .
\end{eqnarray}

It may seem surprising that not a simple difference of one--photon vertices
occurs on the right--hand side of the last equation; however, we stress
(as it will turn out in the next sections)
that this combination of Green's functions and their inverse being
attached to $\Gamma^\nu$ 
is crucial for a correct description of the Compton scattering
process. A similar Ward--Takahasi identity  in the order ${\cal
O}(e^2)$ has also been used in ref. \cite{SchererFearing} in order to
derive the lowest order contributions ({\it i.e.} at order ${\cal
O}(1)$ with respect to the photon momenta) for Compton scattering off
spin--0 particles.

The formulae in eqs. (\ref{WTImu}) and (\ref{WTImunu}) provide general
constraints for the explicit 
one--photon and two--photon vertices $\Gamma^\mu$ and $\Gamma^\munu$
since a violation of these Ward--Takahashi identities leads to
electromagnetic observables that are not gauge invariant; we will show
this in section \ref{GaugeInvariance}.

Before we now insert the explicit expressions for the amputated
Green's functions given in eqs. (\ref{GreensFunction2Order1mu}) and
(\ref{GreensFunction2Order2munu}), we briefly recall the well--known
identities for the free one--fermion propagator

\begin{eqnarray}
\label{OneParticleWTI}
-i(P'-P)_\mu \gamma^\mu =   S^{-1}_F (P') - S^{-1}_F(P)
\end{eqnarray}

and 

\begin{eqnarray}
\label{OneParticleWI}
\gamma^\mu =  i\frac{\partial}{\partial P_\mu} S^{-1}_F (P)
\end{eqnarray}

that --- because $G_0$ includes only free propagators --- immediately
lead to the following relations for the lowest
order vertices $\Gamma^\mu_0$ and $\Gamma^\munu_0$:  

\begin{eqnarray}
\label{FreeWTImu}
-iq_\mu\Gamma_0^\mu(P+q,P) &=& G_0^{-1}(P+q) - G_0^{-1}(P)\\
\mbox{and}\qquad -iq_{1\mu}\Gamma_0^\munu(P+q_1-q_2,P) &=&
G_0^{-1}(P+q_1-q_2)G_0(P-q_2)\Gamma_0^\nu(P-q_2,P) \nonumber\\
\label{FreeWTImunu}
&-& \Gamma_0^\nu(P+q_1-q_2,P+q_1)G_0(P+q_1)G_0^{-1}(P)\quad .
\end{eqnarray}

We stress that due to the definitions in eqs. (\ref{AmplitudesM}), the
Ward--Takahashi identities in this section do not include any charge
factors in physical units. However, these charge factors $e$ and $e^2$,
respectively, can be re--introduced by a simple re--definition of the
$n$--photon irreducible vertices (see section \ref{ChargeFactors});
see also the remarks on charge factors in section
\ref{ExamplesForMesons}. 

Let us finally make some remarks on the apparently different form of
the Ward--Takahashi identities of order ${\cal O}(e)$ and ${\cal O}(e^2)$,
respectively. The difference between the one--photon and the
two--photon irreducible vertex is that $\Gamma^\munu$ in contrast to
$\Gamma^\mu$  includes pole terms. To suppress the contributions from
$\Gamma^\munu_{\mbox{\footnotesize\sc Pole}}$, we have introduced 
$\Gamma^\munu_{\mbox{\footnotesize\sc NoPole}}=\Gamma^\munu -
\Gamma^\munu_{\mbox{\footnotesize\sc Pole}}$ in
eq. (\ref{VertexSplit}). A generalization with and without inclusion
of interaction kernels is given in eq. (\ref{CalGDef}); it is
straightforward to show that these quantities satisfy the following
simple identities (see also \cite{Kazes,SchererFearing}):

\begin{eqnarray}
\label{WTIforCalGmu}
-iq_\mu{\cal G}_{(\epsilon)}^\mu(P+q,P) &=& {\cal G}_{(\epsilon)}(P+q)
- {\cal G}_{(\epsilon)}(P)\\ \mbox{and}\label{WTIforCalGmunu}
-iq_{1\mu}{\cal G}_{(\epsilon)}^\munu(P+q_1-q_2,P) &=& {\cal
G}_{(\epsilon)}^\nu(P-q_2,P) - {\cal G}^\nu_{(\epsilon)}(P', P'+q_2)\quad .
\end{eqnarray}

Both for $\epsilon=1$ and $\epsilon=0$, they can be derived
analogously to the Ward--Takahashi identities given above. We should
note in general that all
identities in order ${\cal O}(e^2)$ can be re--formulated for the
contraction with respect of the four--momentum $q_{2\nu}$ of the second
(outgoing) photon by using crossing symmetry ({\it i.e.}
$q_1\leftrightarrow -q_2, \mu\leftrightarrow\nu$).

\subsection{Identities for the Interaction Kernels}
                                    \label{KernelWTI} 

As we will see in section \ref{GaugeInvariance}, we have to demand
that the Ward--Takahashi identites in eqs. (\ref{WTImu}) and
(\ref{WTImunu}) hold in order to guarantee strict gauge invariance. From
this principle, we can derive analogues identities for the irreducible
interaction kernels. Let us start by inserting the explicit form of
$\Gamma^\mu(P',P)$ into 
the left--hand side of the Ward--Takahashi identity of order ${\cal
O}(e)$:

\begin{eqnarray}
-iq_\mu\Gamma^\mu(P+q,P) &=& -iq_\mu\big(\Gamma_0^\mu(P+q,P)
 -iK^\mu(P+q,P)\big) \nonumber\\
&=& G_0^{-1}(P+q) - G_0^{-1}(P) - q_\mu K^\mu(P+q,P) \nonumber\quad .
\end{eqnarray}

We see from eq. (\ref{WTImu}) that the difference of two inverse full
Green's functions should appear; from this constraint and due to
$G^{-1}=G^{-1}_0+iK$, we can read off a Ward--Takahashi identity for
the one--photon irreducible interaction kernel: 

\begin{eqnarray}
\label{KernelWTImu}
-iq_\mu K^\mu (P+q,P) = K (P) - K(P+q)\quad .
\end{eqnarray}

This result is well--known for a long time and has been derived {\it
e.g.} in refs. \cite{Ohta,GrossRiska}. However, the same procedure can also be
applied to the divergence of the two--photon 
vertex $\Gamma^\munu(P',P)$ with $P'=P+q_1-q_2$ and yields

\begin{eqnarray}
-iq_{1\mu}\Gamma^\munu(P+q_1-q_2,P) &=& -iq_{1\mu}\Gamma_0^\munu(P+q_1-q_2,P)
 -i\big(-iq_{1\mu}K^\munu(P+q_1-q_2,P)\big)\nonumber \\
&&+ \Gamma^\nu(P+q_1-q_2,P+q_1)\:\: G
(P+q_1)\big(-iq_{1\mu}\Gamma^\mu(P+q_1,P)\big) \nonumber
\\&&- \Gamma_0^\nu(P+q_1-q_2,P+q_1) G_0
(P+q_1)\big(-iq_{1\mu}\Gamma_0^\mu(P+q_1,P)\big)\nonumber
\\&&+\big(-iq_{1\mu}\Gamma^\mu (P+q_1-q_2,P-q_2)\big) \:\: G 
(P-q_2)\Gamma^\nu (P-q_2,P) \nonumber
\\&&- \big(-iq_{1\mu}\Gamma_0^\mu (P+q_1-q_2,P-q_2)\big) G_0
(P-q_2)\Gamma_0^\nu (P-q_2,P) \nonumber \\ &=& \:\:\:
G^{-1}(P+q_1-q_2)G(P-q_2)\Gamma^\nu(P-q_2,P) \nonumber\\
&&- \Gamma^\nu(P+q_1-q_2,P+q_1)G(P+q_1)G^{-1}(P) \nonumber\\
&&+ \Gamma_0^\nu(P-q_2,P) - \Gamma^\nu(P-q_2,P)\nonumber\\
&&- \Gamma_0^\nu(P+q_1-q_2,P+q_1) +
\Gamma^\nu(P+q_1-q_2,P+q_1)\nonumber\\
&&- q_{1\mu}K^\munu(P+q_1-q_2,P)\nonumber\qquad ,
\end{eqnarray}

where we have used eqs. (\ref{WTImu}), (\ref{FreeWTImu}) and
(\ref{FreeWTImunu}). By comparison with eq. (\ref{WTImunu})
we observe that the correct terms for the 
Ward--Takahashi identity in ${\cal O}(e^2)$ are already present and
with $\Gamma^\nu_0-\Gamma^\nu=iK^\nu$ find the relation for the
two--photon irreducible interaction kernel:

\begin{eqnarray}
\label{KernelWTImunu}
-iq_{1\mu} K^\munu (P+q_1-q_2,P) = K^\nu (P-q_2,P) - K^\nu(P+q_1-q_2,P+q_1)\quad .
\end{eqnarray}

Note that we have assumed that $K^\nu$ obeys the relation in
eq. (\ref{KernelWTImu}) since we have used the full Ward--Takahashi
identity in the order ${\cal O}(e)$. Let us remark that the identities
in eqs. (\ref{KernelWTImu}) and (\ref{KernelWTImunu}) can be derived
in an even more transparent way by using
eqs. (\ref{KWithCalG})--(\ref{KmunuWithCalG}) together with the
Ward--Takahashi identities for the non--singular quantities ${\cal
G}_{(\epsilon)}^\mu$ and ${\cal G}_{(\epsilon)}^\munu$ in
eqs. (\ref{WTIforCalGmu}) and (\ref{WTIforCalGmunu}).

\subsection{Example: Explicit Identities for $q\bar q$ Mesons}
                                    \label{ExamplesForMesons} 

In eq. (\ref{PeskinWTI}), we have omitted the dependence on the
relative momenta in the amplitudes $M^\mu(P',P)$ and $M(P',P)$; in
general, a sum over all photon couplings to the internal fermion lines and
therefore a dependence of all internal momenta should be written
out. We will demonstrate this for a mesonic system explicitly, {\it
i.e.} for a bound state composed of a quark $q$ with momentum
$p_1$ and charge $e_1$ and an anti--quark  with momentum
$-p_2$ and charge $-e_2$. The two--body Ward--Takahashi identity in
this case reads (see \cite{Bentz}, \cite{Kazes} and references in \cite{Ohta})

\begin{eqnarray}
-i q_\mu M^\mu (p_1 ', -p_2 ';p_1, -p_2) &=& e_1 \big( M(p_1 '-q, -p_2 ';p_1,
-p_2)  - M(p_1 ', -p_2 ';p_1+q, -p_2) \big)  \\
&-& e_2 \big( M(p_1 ', -p_2 '+q;p_1, -p_2) - M(p_1 ', -p_2 ';p_1, -p_2-q) \big)
 \nonumber \quad ;
\end{eqnarray}

note the details in the coupling of the photon to the anti--quark. For
the sake of simplicity, we assume that the operation of the the charge operator $\textsf{Q}_f$ 
is trivial in flavour space and only produces a flavour dependent
factor ({\it e.g. } $e_i=-1/3 e$ or $e_i=+2/3 e$), {\it i.e.} that $\textsf{Q}_f$ is
diagonal in the $f_i=u,d,s$ flavour basis and acts like
$\textsf{Q}_f|f_i\rangle=e_i|f_i\rangle$; see also eq. (\ref{Qquarks})
in section \ref{ChargeFactors}. We shall 
assume here that the interaction 
will not be mediated by charged exchange particles ({\it
i.e.} $[K,\textsf{Q}_f]=0$). Furthermore, we define the total bound state 
momentum by $P=p_1 + p_2$ and the relative momentum by $p=\eta_2 p_1 -
\eta _1 p_2$ (with $\eta_1 + \eta _2 = 1$), see also
eq. (\ref{p1p2_def}). With $P'=P+q$, the two--body Ward--Takahashi
identity in this new coordinates can be written as

\begin{eqnarray}
-i q_\mu M^\mu ( P', p'; P, p ) 
&=& e_1 \big( M( P'-q, p'-\eta _2 q; P, p ) - M( P', p'; P+q, p+\eta _2 q) \big) \\
&-& e_2 \big( M( P'-q, p'+\eta _1 q; P, p ) - M( P', p'; P+q, p-\eta _1 q) \big)
\nonumber \quad .
\end{eqnarray}

Now let us check the free Green's function proportional to the
free amplitude $M_0^\mu=-ieG_0^\mu=-ieG_0 \Gamma^\mu_0 G_0$, explicitly written out in
terms of one--particles propagators:

\begin{eqnarray}
\label{G0GammaGOExplicit}
i M_0^\mu(P',p';P,p) &=& eG_0(P',p')
\left[\Gamma_{0,1}^\mu(P',p';P,p)+\Gamma_{0,2}^\mu(P',p';P,p)\right]G_0(P,p)
\\ 
&=&      \Big[ S_F(\frac{P'}{2}+p')\otimes S_F(-\frac{P'}{2}+p')\Big] \nonumber \\
&\times& \Big( e_1  \cdot\Big[\gamma^\mu \otimes S_F^{-1}(-\frac{P}{2}+p)\Big] \cdot
(2\pi)^8\delta^4(P'-P-q)\delta^4(p'-p-\eta_2 q)\nonumber \\
&-&\:\:\: e_2  \cdot \Big[S_F^{-1}(\frac{P'}{2}+p')\otimes\gamma^\mu\Big]  \cdot
(2\pi)^8\delta^4(P'-P-q)\delta^4(p'-p+\eta_1 q)\Big) \nonumber \\
&\times& \Big[ S_F(\frac{P}{2}+p)\otimes S_F(-\frac{P}{2}+p)\Big]
\nonumber \quad ,
\end{eqnarray}

where $\Gamma_0^\mu$ couples the photon to each fermion
line of the free bound state propagator $G_0$. Using the one--particle
Ward--Takahashi identity in eq. (\ref{OneParticleWTI}), we thus find

\begin{eqnarray}
\lefteqn{-i q_\mu M_0^\mu ( P', p'; P, p )} \\ 
&=& -ie_1 \left( \left[S_F(\frac{P}{2}+p)  \otimes S_F(-\frac{P}{2}+p)\right] 
            - \left[S_F(\frac{P'}{2}+p')\otimes S_F(-\frac{P}{2}+p)\right] \right) \cdot
(2\pi)^8\delta^4(P'-P-q)\delta^4(p'-p-\eta_2 q) \nonumber\\
&& +ie_2 \left( \left[S_F(\frac{P}{2}+p)  \otimes S_F(-\frac{P}{2}+p)\right] 
            - \left[S_F(\frac{P}{2}+p)\otimes S_F(-\frac{P'}{2}+p')\right] \right) \cdot
(2\pi)^8\delta^4(P'-P-q)\delta^4(p'-p+\eta_1 q)\nonumber
\end{eqnarray}

We now define $\tilde e_i := e_i / e$ and write down the related
expressions for the amputated one--photon irreducible vertices
$e\Gamma^\mu = iG^{-1} M^\mu G^{-1}$ and $e\Gamma_0^\mu = iG_0^{-1}
M_0^\mu G_0^{-1}$: 

\begin{eqnarray}
-i q_\mu \Gamma^\mu ( P', p'; P, p ) 
&=& \tilde e_1 \big( G^{-1}( P', p'; P+q, p+\eta _2 q) - G^{-1}( P'-q, p'-\eta _2 q; P, p )\big) \\
&-& \tilde e_2 \big( G^{-1}( P', p'; P+q, p-\eta _1 q) - G^{-1}( P'-q, p'+\eta _1 q; P, p )\big)
\nonumber 
\end{eqnarray}

and 

\begin{eqnarray}
-i q_\mu \Gamma_0^\mu ( P', p'; P, p ) 
&=& \tilde e_1 \big( \hat G_0^{-1}( P', p'; P+q, p+\eta _2 q) - \hat
G_0^{-1}( P'-q, p'-\eta _2 q; P, p )\big) \\ 
&-& \tilde e_2 \big( \hat G_0^{-1}( P', p'; P+q, p-\eta _1 q) - \hat
G_0^{-1}( P'-q, p'+\eta _1 q; P, p )\big) 
\nonumber \quad ;
\end{eqnarray}

here, the inverse free propagator $\hat G_0$ (depending on primed
and non--primed coordinates) is defined by 

\begin{eqnarray}
\hat G_0^{-1}( P', p'; P, p) := \left[S_F^{-1}(\frac{P}{2}+p)\otimes
S_F^{-1}(-\frac{P}{2}+p)\right]\cdot 
(2\pi)^8\delta^4(P'-P)\delta^4(p'-p) \quad .
\end{eqnarray}

If we recall $G^{-1}=\hat G^{-1}_0+iK$ as well as the general relation
$\Gamma^\mu = \Gamma^\mu_0 - iK^\mu$ for the one--photon irreducible
vertex, then we easily find with $P'=P+q$

\begin{eqnarray}
\label{KernelWTImuExplicit}
-i q_\mu K^\mu ( P', p'; P, p ) 
&=& \tilde e_1 \Big( K( P'-q, p'-\eta _2 q; P, p ) - K( P', p'; P+q, p+\eta _2 q) \Big) \\
&-& \tilde e_2 \Big( K( P'-q, p'+\eta _1 q; P, p ) - K( P', p'; P+q, p-\eta _1 q) \Big)
\nonumber \quad .
\end{eqnarray}

This is the full version of the constraint in
eq. (\ref{KernelWTImu}), formulated explicitly for a $q\bar q$
system without omitting the relative momenta. As we have pointed out
in the last subsection, the divergence of $K^\mu$ vanishes for $q\to
0$. We now see in detail that this also holds if the interaction
kernel $K$ is independent of the total four--momentum $P$ and is
simultaneously of local type; then the terms on the right--hand
side of eq. (\ref{KernelWTImuExplicit}) mutually cancel if
$K\sim K(p'-p)$ is a convolution--like kernel.

In analogy to this derivation, we find the explicit formulation of the
Ward--Takahashi identity for the two--photon irreducible interaction
kernel of a $q\bar q$ system with $P+q_1=P'+q_2$:

\begin{eqnarray}
\label{KernelWTImunuExplicit}
-i q_{1\mu} K^\munu ( P', p'; P, p ) 
&= \tilde e_1 &\Big( K^\nu( P'-q_1, p'-\eta _2 q_1; P, p ) - K^\nu(
P', p'; P+q_1, p+\eta _2 q_1) \Big) \\ 
&- \tilde e_2& \Big( K^\nu( P'-q_1, p'+\eta _1 q_1; P, p ) - K^\nu(
P', p'; P+q_1, p-\eta _1 q_1) \Big)
\nonumber   \quad .
\end{eqnarray}

Note that for a deuteron described as a $pn$ bound state ({\it i.e.}
with $\tilde e_1 = \tilde e_p = 1$ and $\tilde e_2 = \tilde e_n = 0$),
the r.h.s. of this equation reduces to the difference in the first line. The resulting
identity (up to a factor $ie$ and in a slightly different notation)
for the relative momentum parameter choice $\eta_1=\eta_2=\frac 1 2 $ is completely
equivalent to eq. (4.13) in ref. \cite{GrossIto}; in this publication,
the authors have found this constraint for the two--photon interaction
current by demanding gauge invariance of the Compton scattering
amplitude in a general Bethe--Salpeter framework. In fact, this
represents a completely different approach to this issue compared to the one shown
here. While we find the Ward--Takahashi identities for the $n$--photon
irreducible interaction kernels in a rather general framework starting
from the ``gauging of equations method'' applied to the Green's
function of a composite system, the authors of ref. \cite{GrossIto}
begin with the description of a special bound state ({\it i.e.} the
deuteron system); however, they keep a certain universality since they
do not specify the interaction kernel in the Bethe--Salpeter
equation. Then they consider Compton scattering off this system,
include the lowest order contribution ($\sim \Gamma^\munu_0$),
re--scattering terms via $T$ matrix diagrams and one--photon
interaction kernels $K^\mu$, and study the gauge invariance of the
resulting tensor. They find that $-i q_{1\mu} T^\munu=0$ only holds if
an additional two--photon irreducible term is included; this condition
is then recasted in the condition quoted in
eq. (\ref{KernelWTImunuExplicit}). 

The direct derivation of the identity in
eq. (\ref{KernelWTImunuExplicit}) is lengthy since for instance
the (explicitly crossing--symmetric) subtraction terms in
eq. (\ref{GreensFunction2Order2munu}) for a $q\bar q$ system read
explicitly

\begin{eqnarray}
\label{GmuExplicit}
\lefteqn{ \left.\Big[\Gamma_0^\nu(P',P'+q_2) G_0
(P'+q_2)\Gamma_0^\mu(P+q_1,P) + \Gamma_0^\mu(P',P'-q_1) G_0
(P'-q_1)\Gamma_0^\nu(P-q_2,P)\Big]\right | _{q\bar q}} \\
&\hat = &\:\:\:\tilde e_1^2 \:\:\:
\Gamma_{0,1}^\nu(P',p'+\eta _2 q;P'+q_2, p''+\eta _2 q_1)
G_0(P'+q_2,p''+\eta _2 q_1;P+q_1, p'''+\eta _2 q_1)
\Gamma_{0,1}^\mu(P+q_1,p'''+\eta _2 q_1;P, p) \nonumber \\
&+ &\:\:\:\tilde e_2^2  \:\:\:
\Gamma_{0,2}^\nu(P',p'-\eta _1 q;P'+q_2, p''-\eta _1 q_1)
G_0(P'+q_2,p''-\eta _1 q_1;P+q_1, p'''-\eta _1 q_1)
\Gamma_{0,2}^\mu(P+q_1,p'''-\eta _1 q_1;P, p)\nonumber \\
&- &\:\tilde e_1 \tilde e_2\: 
\Gamma_{0,2}^\nu(P',p'+\:\,Q\,\:;P'+q_2, p''+\eta _2 q_1)
G_0(P'+q_2,p''+\eta _2 q_1;P+q_1, p'''+\eta _2 q_1)
\Gamma_{0,1}^\mu(P+q_1,p'''+\eta _2 q_1;P, p)\nonumber \\
&- &\:\tilde e_1 \tilde e_2 \:
\Gamma_{0,1}^\nu(P',p'-\:\,Q\,\:;P'+q_2, p''-\eta _1 q_1)
G_0(P'+q_2,p''-\eta _1 q_1;P+q_1, p'''-\eta _1 q_1)
\Gamma_{0,2}^\mu(P+q_1,p'''-\eta _1 q_1;P, p)\nonumber \\
&+& \left\{ 
\begin{array}{ccc} 
q_1& \leftrightarrow &-q_2\\
\mu& \leftrightarrow & \nu
\end{array}\right\} \nonumber \quad ,
\end{eqnarray}

where we have defined $q:=q_1 - q_2$ and $Q:= \eta _2 q_1 + \eta _1
q_2$; the definition of $\Gamma^\mu_{0,i}$ can be read off
eq. (\ref{G0GammaGOExplicit}). Note that we still neither
indicate the integrations $\int d^4p''$ and $\int d^4p'''$ nor the
various indices; see eqs. (\ref{GreensFunctionEquation}) and
(\ref{GreensFunctionEquationExplicit}) for comparison. Due to the 
complexity and the large number of 
terms occuring in the explicit $q\bar q$ formulation of the
two--photon irreducible vertex $\Gamma^\munu$, we feel that it might
be more instructive to keep on omitting the relative momenta and the
terms related to internal charges in the rest of this contribution
since it will clarify the structure of the resulting expressions.
We stress however that from contributions like those quoted above
it is in fact straightforward to derive the explicit Ward--Takahashi
identities for a $q\bar q$ system in the order ${\cal O}(e^2)$. 

We have chosen the explicit example in eq. (\ref{GmuExplicit}) not
only to justify the suppression of relative momenta in this paper but
also to illustrate the role of the subtraction terms $-\Gamma^\mu_0
G_0\Gamma^\nu_0-\Gamma^\nu_0 G_0\Gamma^\mu_0$. Note that if we write
out the crossing symmetric terms in this example by exchanging $q_1
\leftrightarrow -q_2$ and $\mu \leftrightarrow  \nu$ and by explicitly
using the one--particle propagator formulation as in
eq. (\ref{G0GammaGOExplicit}), we find that the 
$\tilde e_i^2$ parts are correctly produced but that a
doubling of terms proportional to $\tilde e_1 \tilde e_2$ occurs; these
terms are essentially proportional to $\gamma ^\mu\otimes\gamma^\nu$
and $\gamma ^\nu\otimes\gamma^\mu$, respectively. This
is precisely what happens in the lowest--order parts of $+\Gamma^\mu
G\Gamma^\nu+\Gamma^\nu G\Gamma^\mu$ --- only with an opposite sign. By
this mechanism, it is guaranteed that the only lowest--order terms
surviving in the complete two--photon irreducible vertex
$\Gamma^\munu$ are collected in $\Gamma^\munu_0$ without any
double--counting.

\subsection{Identities for the Bethe--Salpeter Amplitudes}
                                    \label{BSAWTI} 

Now that we have derived the Ward--Takahashi identities for the
$n$--photon vertices and irreducible interaction kernels, we are able
to give similar 
relations for the ``extended'' Bethe--Salpeter amplitudes of orders
${\cal O}(e)$ and ${\cal O}(e^2)$.

If we take the divergence of eq. (\ref{BSAmu}), then we find
immediately with the Ward--Takahashi identity for the one--photon
vertex the relation

\begin{eqnarray}
\label{ChiWTImu}
-iq_{\mu} \chi^\mu (P+q,P) = \chi(P) \qquad ;
\end{eqnarray}

here, we have used the Bethe--Salpeter equation
$G^{-1}(P)\chi(P)=0$. Analogously, the relation for the second order
Bethe--Salpeter amplitude reads

\begin{eqnarray}
\label{ChiWTImunu}
-iq_{1\mu} \chi^\munu (P+q_1-q_2,P) = \chi ^\nu (P-q_2,P) \quad .
\end{eqnarray}

As we have already stated in section \ref{BetheSalpeterAmplitudes}, the
relations above are well defined for finite photon momenta.

\subsection{Differential Identities}
                                    \label{DiffWTI} 

We will now consider differential forms of the
relations in the preceeding subsections. The limit $q, q_1, q_2\to
0$ yields $P'\to P$; therefore we suppress all momentum dependencies
in the following.

With $\partial^\mu:=\partial/\partial P_\mu$ or equivalently assuming
infinitesimally small photon momenta, we find for the extended
Green's functions 

\begin{eqnarray}
\label{delG}
\partial^\mu G &=& i G^\mu \\
\label{delGmu}
\mbox{and} \qquad\partial^\nu G^\mu &=& i G^\munu \quad .
\end{eqnarray}

From eqs. (\ref{KernelWTImu}) and (\ref{KernelWTImunu}), we obtain the
following relations for the one--photon and two--photon irreducible
interaction kernels:

\begin{eqnarray}
\partial^\mu K &=& i K^\mu \\
\mbox{and} \qquad\partial^\nu K^\mu &=& i K^\munu \quad .
\end{eqnarray}

In order ${\cal O}(e)$, the situation for the amputated Green's
function is simple. However, in the second order relation in
eq. (\ref{WTImunu}) no simple difference with shifted arguments occurs
on the right--hand side. To derive a Ward identity for $\Gamma^\munu$,
we therefore start with eq. (\ref{delGmu}) and recall that
$G^\munu=G\Gamma^\munu G$:

\begin{eqnarray}
\partial^\nu G^\mu &=& \partial^\nu (G\Gamma^\mu G) \nonumber\\
&=& (\partial^\nu G)\Gamma^\mu G+G(\partial^\nu\Gamma^\mu)
G+G\Gamma^\mu (\partial^\nu G) \nonumber \\
&=& (iG^\nu)\Gamma^\mu G+G(\partial^\nu\Gamma^\mu)
G+G\Gamma^\mu (iG^\nu) \nonumber \\
&=& G(i\Gamma^\nu G\Gamma^\mu +i\Gamma^\mu G\Gamma^\nu +
\partial^\nu\Gamma^\mu ) G \nonumber \\
&\stackrel{!}{=}&iG\Gamma^\munu G\quad .\nonumber
\end{eqnarray}

For the one--photon and two--photon vertices, the corresponding
relations with $\Gamma:=G^{-1}$ are therefore

\begin{eqnarray}
\label{delGamma1stOrder}
\partial^\mu \Gamma  &=& -i \Gamma ^\mu \\
\label{delGamma2ndOrder}
\mbox{and} \qquad\partial^\nu \Gamma ^\mu &=& -i (\Gamma^\nu G\Gamma
^\mu + \Gamma^\mu G\Gamma ^\nu -\Gamma ^\munu) \quad .
\end{eqnarray}

Again we find that the differential two--photon relation for the vertex is not a
simple analogue to the one--photon case; however, both orders of the
electromagnetic coupling have in common that exclusively non--singular
terms appear since we can write $\partial^\nu \Gamma ^\mu = i
\Gamma^\munu_{\mbox{\footnotesize\sc NoPole}}$ for the last identity,
see eq. (\ref{VertexSplit}). We will see in section
\ref{LowEnergyLimit} that the non--singularity of the last equation
will be crucial for the correct low--energy limit of the Compton
scattering tensor; note that the sign in eq. (\ref{delGamma2ndOrder})
is essential for the related cancellation in the limit $q_i\to 0$.

For the sake of completeness, we also quote the analoguous results with
$\Gamma_0:=G_0^{-1}$ for the free vertices:

\begin{eqnarray}
\label{delFreeGamma1stOrder}
\partial^\mu \Gamma_0  &=& -i \Gamma_0 ^\mu \\
\label{delFreeGamma2ndOrder}
\mbox{and} \qquad\partial^\nu \Gamma_0 ^\mu &=& -i ( \Gamma_0^\nu
G_0\Gamma_0 ^\mu  + \Gamma_0^\mu G_0\Gamma_0 ^\nu -\Gamma_0 ^\munu)\quad .
\end{eqnarray}

We will finish this section by studying differential identities for
$\chi$, $\chi^\mu$ and $\chi^\munu$ comparable to those above. Let us
anticipate that the following relations basing on eqs. (\ref{BSAmu})
and (\ref{BSAmunu}) are derived in a quite formal way, {\it i.e.} 
regardless of the poles that will appear in the Green's function
$G(\tilde P)$ for $\tilde P\to P$ ($P^2=M^2$) or equivalently for
vanishing photon momenta 

To find the derivative of the Bethe--Salpeter amplitudes, we cannot
start from the relation in eq. (\ref{ChiWTImu}) since there no
difference appears. However, we can exploit the Bethe--Salpeter
equation and find

\begin{eqnarray}
\label{delChi}
\partial ^\mu \chi &=& \partial ^\mu (-iG_0K\chi) =
-i\big((\partial ^\mu G_0)K\chi + G_0(\partial ^\mu K)\chi + G_0
K(\partial ^\mu\chi)\big) \\ 
&=& \underbrace{(\Id+iG_0 K)^{-1}}_{=GG_0^{-1}} G_0\big(\Gamma _0^\mu \underbrace{G_0
K\chi}_{=i\chi} + K^\mu\chi\big) \nonumber\\
&=& iG(\Gamma_0^\mu - iK^\mu)\chi = iG\Gamma^\mu \chi\nonumber \qquad .
\end{eqnarray}

The second derivative of $\chi$ can be calculated analogously by using
eq. (\ref{delGamma2ndOrder}) such that characteristic cancellations occur;
we find with eqs. (\ref{BSAmu}) and (\ref{BSAmunu}) the following relations:

\begin{eqnarray}
\label{delChi1stOrder}
\partial ^\mu \chi &=& i \chi^\mu \;= iG\Gamma^\mu \chi \\
\label{delChi2ndOrder}
\mbox{and} \qquad\partial ^\nu \chi ^\mu&=& i \chi^\munu   = iG\Gamma^\munu \chi \quad .
\end{eqnarray}

Similar equations hold for the adjoint amplitudes:

\begin{eqnarray}
\label{delbarChi1stOrder}
\partial ^\mu \bar\chi &=& i \bar\chi^\mu \;= i\bar\chi\Gamma^\mu G \\
\label{delbarChi2ndOrder}
\mbox{and} \qquad\partial ^\nu \bar\chi ^\mu&=& i \bar\chi^\munu   =
i\bar\chi\Gamma^\munu G  \quad . 
\end{eqnarray}

Note that the bound state Bethe--Salpeter amplitudes are defined only for
on--shell momenta $P^2=M^2$ whereas the Green's function $G(\tilde P)$
generically includes off--shell contributions. Therefore the
above--mentioned equations actually have to be considered with care:
any derivative of $\chi$ would infinitesimally shift the Bethe--Salpeter amplitude
away from mass--shell in a region where $\chi$ becomes
ill--defined. However, we can formally introduce a quantity
${\cal X}(\tilde P)$ for off--shell momenta $\tilde P^2\not =M^2$
analogously to the definition of $\chi$ in eq. (\ref{DefBSA}); then ${\cal X}(\tilde P)$ is not
a solution of the Bethe--Salpeter equation but we can properly define the
operation $\partial ^\mu{\cal X}$. Additional terms due to
$G^{-1}{\cal X}\not =0$ will appear in eq. (\ref{delChi}) that
vanish in the limit $\tilde P^2\to M^2$ in which we find the final
relations in eqs. (\ref{delChi1stOrder})--(\ref{delbarChi2ndOrder}).

We shall remark here that the Green's function $G$ that enters in
eqs. (\ref{delChi1stOrder})--(\ref{delbarChi2ndOrder}) exhibits a pole
in the on--shell case and (for 
states with total momentum $\tilde P$ and mass $M$) can be
written (see {\it e.g.} \cite{Lurie}) as 

\begin{eqnarray}
\label{GreensPole}
 G(\tilde P) = -i\: \frac{\chi(P)\circ\bar\chi(P)}{\tilde P^2 - M^2}\quad +\quad
 {\mbox{terms regular}\atop\mbox{for}\quad \tilde P^2\to M^2}\quad .
\end{eqnarray}

Note that the Bethe--Salpeter amplitudes are on--shell quantities and
therefore depend on $P$ with $P^2=M^2$. The physical implication of
the dominant term in the Green's function $G(\tilde P)$ for the derivative of 
the Bethe--Salpeter amplitude and the related first order expanded
amplitude $\chi^\mu$ in the limit $\tilde P^2\to M^2$ is depicted diagrammatically in
fig. (\ref{fig:delChi1stOrder}); an analoguous diagram can be sketched
for the second order Bethe--Salpeter amplitude $\chi^\munu$.

\protect\begin{figure}[h]
  \protect\begin{center}
    \leavevmode
       \protect\input{delChi1.pstex_t}
\newline
       \protect\caption{The Bethe--Salpeter amplitude expanded in
       first order of the electromagnetic field according to
       eq. (\protect\ref{delChi1stOrder}).} 
       \protect\label{fig:delChi1stOrder}
   \end{center}
\end{figure}

We stress again that these differential relations for the
Bethe--Salpeter amplitudes $\chi$ and $\bar\chi$ are derived with
quite formal arguments and without 
discussing the apparent poles that enter the equations above. However,
they allow for an alternative derivation of 
low--energy limit for the Compton scattering process as we shall show
in section \ref{BornThomsonLimit}.

\section{Gauge Invariance}        
                                        \label{GaugeInvariance}

It is an important question whether the one--photon and two--photon
vertices defined in the eqs. (\ref{GreensFunction2Order1mu}) and
(\ref{GreensFunction2Order2munu}) satisfy gauge invariance if we
evaluate them between Bethe--Salpeter amplitudes. To
formulate this task more precisely, let us first define the
electromagnetic current via

\begin{eqnarray}
\label{CurrentDef}
J^\mu := \langle P' \left| j^\mu (0) \right|P\rangle = - \bar\chi(P')\Gamma^\mu(P',P)\chi(P)
\end{eqnarray}

and the Compton scattering tensor via

\begin{eqnarray}
\label{DefTmunu}
T^\munu := i\int d^4 x \langle P' \left| \textsf{T}j^\mu (x) j^\nu
(0)\right|P\rangle e^{-iq_1 x} =  i \bar\chi(P')\Gamma^\munu(P',P)\chi(P)\quad .
\end{eqnarray}

Here, $\textsf{T}$ denotes the time--ordered product. Gauge invariance
implies that the divergences of $J^\mu$ and $T^\munu$ vanish, {\it i.e.}
that in momentum space

\begin{eqnarray}
-iq_\mu J^\mu  = 0 \qquad \mbox{and}\qquad -iq_{1\mu}T^\munu =
 -iT^\munu q_{2\nu} = 0\nonumber
\end{eqnarray}

holds. In the subsequent paragraphs, we will address this question for
one--photon and two--photon amputated Green's functions in various
approximations. Let us furthermore stress that in the following two
sections the momenta $P$ and $P'$ of the incoming and outgoing bound
state are supposed to be on--shell ({\it i.e.} $P^2=P'^2=M^2$)  since
otherwise the Bethe--Salpeter equations could not be applied.

\subsection{Electromagnetic Current}
                                    \label{Jmu} 

With the Ward--Takahashi identity for the one--photon vertex in
eq. (\ref{WTImu}), it is easy to show that the divergence of the
electromagnetic current vanishes (here, energy--momentum conservation
requires $P'=P+q$):

\begin{eqnarray}
-iq_\mu J^\mu  &=& -  \bar\chi(P')\big(-iq_\mu
 \Gamma^\mu(P',P)\big)\chi(P)\nonumber\\
&=& -  \bar\chi(P')\big(G^{-1}(P')-G^{-1}(P)\big)\chi(P) = 0 \nonumber\quad .
\end{eqnarray}

Obviously it is trivial to show the gauge invariance of the
electromagnetic current with the help of the Bethe--Salpeter equations
$ G^{-1}(P)\chi(P)= \bar\chi(P')G^{-1}(P') = 0$. However, this
requires that the one--photon vertex obeys the Ward--Takahashi
identity in eq. (\ref{WTImu}) which is only true if the
one--photon irreducible interaction kernel $K^\mu$ satisfies the
relation in eq. (\ref{KernelWTImu}). To see what happens if we neglect
contributions of this one--photon kernel, 
{\it i.e.} if we only consider the lowest order vertex $\Gamma
_0^\mu$, we again exploit the Bethe--Salpeter equations
$ G_0^{-1}(P)\chi(P) = -i K(P)\chi(P)$ and $\bar\chi
(P')G_0^{-1}(P') = - i\bar\chi(P') K(P') $ and find 

\begin{eqnarray}
\label{GaugeInvLowOrderJ}
-iq_\mu J_0^\mu  &=& -  \bar\chi(P')\big(-iq_\mu \Gamma_0^\mu(P',P)\big)\chi(P)\\ 
&=& -  \bar\chi(P')\big(G_0^{-1}(P')-G_0^{-1}(P)\big)\chi(P) \nonumber\\
&=& i  \bar\chi(P')\big(K(P')-K(P)\big)\chi(P)\nonumber\quad .
\end{eqnarray}

This expression only vanishes if the interaction kernel does not
depend on the bound states' four--momenta. Furthermore it turns out that the kernel has
to be of local type --- {\it i.e.} $K(P;\{k_j\}, \{k_j'\})= K(\{k_j -
k_j'\})$ --- if we explicitly write out the dependence on the relative
momenta. This is true for kernels in ladder approximation but not
necessarily for instantaneously approximated kernels (as {\it
e.g.} used in \cite{KollRicken1,KollRicken2}). 

Note that, independent of the kernel type, we can state that
$-iq_\mu J_0^\mu$ vanishes and therefore gauge invariance holds for
$q= P'-P\to 0$.

\subsection{Compton Scattering Tensor}
                                    \label{Tmunu} 

As for the electromagnetic current, it is again easy to show that the
divergence of the Compton scattering tensor vanishes if we take into
account all contributions in $\Gamma^\munu (P',P)$. Then the
Ward--Takahashi identity for the two--photon vertex in
eq. (\ref{WTImunu}) holds if the included $n$--photon irreducible interaction
kernels obey eqs. (\ref{KernelWTImu}) and (\ref{KernelWTImunu}). With the Bethe--Salpeter
equations, we therefore find for $P'=P+q_1-q_2$

\begin{eqnarray}
-iq_{1\mu} T^\munu  &=&  i \bar\chi(P')\big(-iq_{1\mu}
\Gamma^\munu(P',P)\big)\chi(P)\nonumber \\
&=&  i
\bar\chi(P')\big(G^{-1}(P')G(P-q_2)\Gamma^\nu(P-q_2,P)-\Gamma^\nu(P',P+q_1)G(P+q_1)
G^{-1}(P)\big)\chi(P) = 0 \nonumber\quad .
\end{eqnarray}

As an approximation, we now want to neglect the contributions from
one--photon and two--photon irreducible interaction kernels. Then the
subtraction term in eq. (\ref{GreensFunction2Order2munu}) cancels the
lowest order part in $\Gamma^\gamma G \Gamma^\gamma$ --- see
eq. (\ref{HatGreensFunction2Order2}) --- and we start with
a modified two--photon vertex reading explicitly 

\begin{eqnarray}
\label{ModifiedGammamunu}
\hat\Gamma^\munu (P',P) = \Gamma_0^\munu(P',P) &-& i\Gamma_0^\nu(P',P'+q_2) T
(P'+q_2)\Gamma_0^\mu(P+q_1,P) \nonumber\\ &-& i\Gamma_0^\mu(P',P'-q_1) T
(P\: -q_2)\Gamma_0^\nu(P-q_2,P) \qquad ,
\end{eqnarray}

where the $T$ matrix satisfies eq. (\ref{TMatrixEquation}). The
divergence of this vertex can easily be derived from the
Ward--Takahashi identities for the free one--photon and two--photon
vertices in eqs. (\ref{FreeWTImu}) and (\ref{FreeWTImunu}):

\begin{eqnarray}
-iq_{1\mu}\hat\Gamma^\munu (P',P) &=& G_0^{-1}(P+q_1-q_2)G_0(P-q_2)\Gamma_0^\nu(P-q_2,P)\nonumber \\
&-& \Gamma_0^\nu(P+q_1-q_2,P+q_1)G_0(P+q_1)G_0^{-1}(P) \nonumber \\
&-& i\Gamma_0^\nu(P+q_1-q_2,P+q_1)T(P+q_1)\Big(G_0^{-1}(P+q_1)-G_0^{-1}(P)\Big) \nonumber \\
&-& i\Big(G_0^{-1}(P+q_1-q_2)-G_0^{-1}(P-q_2)\Big)T(P-q_2)\Gamma_0^\nu(P-q_2,P)\nonumber\quad .
\end{eqnarray}

Sandwiching this expression between a Bethe--Salpeter amplitude and
its adjoint, using $ G_0^{-1}\chi = -i K\chi$ and $\bar\chi G_0^{-1} =
- i\bar\chi K$, respectively, and inserting the $T$ matrix equation
--- see fig. (\ref{fig:TMatrixEquation}) ---,
we obtain the divergence of the Compton scattering tensor as

\begin{eqnarray}
-iq_{1\mu}\hat T^\munu  &=&   i \bar\chi(P')\Big(-iq_{1\mu}\hat
 \Gamma^\munu(P',P)\Big)\chi(P)\nonumber  \\
 &=&   i \bar\chi(P')\Big((-iK(P')) G_0(P-q_2) -
 i(-iK(P'))T(P-q_2)\nonumber\\ &&\mbox{\hspace*{4.3cm}} +
 i G_0^{-1}(P-q_2)T(P-q_2)\Big)\Gamma^\nu_0(P-q_2,P)\chi(P)\nonumber\\
 &&  - i \bar\chi(P')\Gamma^\nu_0(P',P+q_1)\Big( G_0(P+q_1)(-iK(P)) -
 iT(P+q_1)(-iK(P)) \nonumber\\  &&\mbox{\hspace*{4.3cm}}+
 i T(P+q_1)G_0^{-1}(P+q_1)\Big)\chi(P)\nonumber\\
 &=&    \bar\chi(P')\Big(K(P') G_0(P-q_2) - iK(P')T(P-q_2) -
 K(P-q_2)G_0(P-q_2)\nonumber\\ && \mbox{\hspace*{4.3cm}}
 +iK(P-q_2)T(P-q_2)\Big)\Gamma^\nu_0(P-q_2,P)\chi(P)\nonumber\\ 
 &&  -  \bar\chi(P')\Gamma^\nu_0(P',P+q_1)\Big( G_0(P+q_1)K(P) - iT(P+q_1)K(P) -
  G_0(P+q_1)K(P+q_1)\nonumber\\ && \mbox{\hspace*{4.3cm}}
 +iT(P+q_1)K(P+q_1)\Big)\chi(P)\nonumber\\ 
 &=&    \bar\chi(P')\Big(K(P') - K(P-q_2)\Big) \Big(G_0(P-q_2) -
 iT(P-q_2)\Big)\Gamma^\nu_0(P-q_2,P)\chi(P)\nonumber\\ 
 &+&    \bar\chi(P')\Gamma^\nu_0(P',P+q_1)\Big( G_0(P+q_1) -
 iT(P+q_1)\Big)\Big(K(P+q_1)-K(P)\Big)\chi(P)\nonumber\quad .
\end{eqnarray}

With $G=G_0-iT$ and defining $\hat\chi^\mu:=G\hat\Gamma^\mu\chi$
(where $\hat\Gamma^\mu=\Gamma_0^\mu$) in analogy to
eqs. (\ref{BSA1stOrder}) and (\ref{BSA2ndOrder}), we can therefore write
the divergence of Compton scattering 
tensor without inclusion of $n$--photon irreducible interaction
kernels as

\begin{eqnarray}
-iq_{1\mu}\hat T^\munu  &=&    \bar{\chi}(P')\Big(K(P') - K(P'-q_1)\Big)
\hat{\chi}^\nu(P-q_2,P)\nonumber\\  
 &+&   \bar{\hat{\chi}}^\nu(P',P'+q_2)
 \Big(K(P+q_1)-K(P)\Big)\chi(P)\nonumber  \quad .  
\end{eqnarray}

This expression vanishes --- as in the similar case of the
lowest--order electromagnetic current, see eq. (\ref{GaugeInvLowOrderJ}) ---
in the limit $q_1, q_2\to 0$ ($P'\to P$). The gauge invariance
condition $-iq_{1\mu}\hat T^\munu=0$ also holds if we use an interaction
kernel $K(P)$ that does not depend on the total four--momenta of the
bound states. As it has been pointed out in the preceeding subsection
for the lowest order current $J_0^\mu$,
an inclusion of all relative momenta in the calculation shows that the
kernel must also be local in order to satisfy gauge invariance in this case. 
This result has also been found in \cite{GrossIto} where an
explicit calculation for deuteron Compton scattering has been
presented. We agree completely with the authors who stress that this
observation emphasizes the necessity of the inclusion of $n$--photon
irreducible interaction kernels if the constituents of the composite
system interact via non--local potentials.

The last point to study is the question of gauge invariance for the
lowest order two--photon vertex $\Gamma _0^\munu$. If we recall the
free Ward--Takahashi identity for this vertex in
eq. (\ref{FreeWTImunu}), we find immediately

\begin{eqnarray}
\label{GaugeInvariance0}
-iq_{1\mu} T_0^\munu  &=&  i \bar\chi(P')\big(-iq_{1\mu}
\Gamma_0^\munu(P',P)\big)\chi(P)\nonumber  \\
&=& i\bar\chi(P')\big(G_0^{-1}(P')G_0(P-q_2)
\Gamma_0^\nu(P-q_2,P)-\Gamma_0^\nu(P',P'+q_1)G_0(P+q_1)  
G_0^{-1}(P)\big)\chi(P) \nonumber  \quad .
\end{eqnarray}

Without applying the Bethe--Salpeter equation, we see that this lowest
order expression only vanishes for $q_1, q_2\to 0$. It is clear that
in this limit the re--scattering terms due to the $T$ matrix will not
contribute; therefore gauge invariance in the very--low--energy regime
will be approximately restored even without explicit inclusion of
intermediate states.

\section{Low--Energy Limits}        
                                        \label{LowEnergyLimit}

In this section, we will study the low--energy behaviour of the
one--photon expression $\bar\chi\Gamma^\mu\chi$ and the two--photon
expression $\bar\chi\Gamma^\munu\chi$. We will restrict our
discussions to bound states with total angular momentum $J=0$ so that
we will not have to consider technical subtleties originating from non--zero
spins; however, our statements remain valid also for spin--averaged
amplitudes describing Compton scattering {\it e.g.} off a proton or a
neutron with total spin $\frac 1 2$. 

In the first subsection, we will give a concise formalism how to include
the correct charge factors. To clearify this point, we will pin down
the lowest order results for the electromagnetic current and the Compton scattering
tensor in explicit expressions for the bound state being a
(pseudoscalar or scalar) $q\bar q$ meson as an example.

\subsection{Charge Factors}
                                    \label{ChargeFactors} 

If we now study the low--energy limits of the electromagnetic current
and the Compton scattering tensor, a caveat is
in  order at this point. Up to now, we have considered the
photon coupling in Dirac space only, which was of the (correct) vector type
$\gamma^\mu$; we thus neglected the (in the general case iso--spin
dependent) charge operator $\textsf{Q}_f$ in 
flavour space that reads {\it e.g.} for hadronic bound states 

\begin{eqnarray}
\label{Qquarks}
\textsf{Q}_f = e\left(
\begin{array}{ccc}
\frac 2 3 & 0 & 0 \\
0& -\frac 1 3& 0 \\
0 & 0 & -\frac 1 3 \\
\end{array}
\right)
\end{eqnarray}

for constituents being $u$, $d$ and $s$ quarks (see also
\cite{Blankleider2,Ohta} for a different formulation of this
issue). This observation does 
not spoil our results on Ward--Takahashi identities and gauge
invariance but it turns out in this section that it will become
important for the low--energy limits. Instead of the
operator in eq. (\ref{Qquarks}), we simply get $\textsf{Q}_f =\Id$ if we introduce our vector
coupling in Dirac space by a derivative of the one--particle
propagators. However, these fermion propagators $S_F(p_i)$ are functions of the
momentum of the $i$--th constituent which depends in a simple
way of the total momentum $P$ (see appendix
\ref{app:Coordinates}). This dependence is given in
eq. (\ref{pPRelation}) and yields

\begin{eqnarray}
\frac{\partial}{\partial P_\mu} S_F(p_j) = -i \eta _j  S_F(p_j) \gamma
^\mu S_F(p_j)\quad .\nonumber 
\end{eqnarray}

As described in the appendix, the $\eta _j$'s are parameters that set
a special choice of relative momenta; we can fix them by $\eta _j=e_j/{\cal Q}$
with ${\cal Q}=\sum _j e_j$ being the total charge and $e_j$ being the
charge of the $j$--th constituent. 

We now have to specify what we mean in detail with the lowest order one--photon
vertex $\Gamma_0^\mu$ for a composite system of $N$
constituents. Recalling $G_0^{-1}=:\Gamma_0$ where the total momentum
is unchanged by $G_0$ ({\it i.e.} $P=\sum _j p_j=\sum _j p_j^\prime$), we state with
eq. (\ref{delFreeGamma1stOrder}) that

\begin{eqnarray}
\label{GammaMuExplicitly}
i\frac{\partial}{\partial P_\mu} G_0^{-1}(P, \{k_i^\prime\}, \{k_i\})
&=& i\frac{\partial}{\partial P_\mu} G_0^{-1}(\{p_j^\prime\}, \{p_j\}) \nonumber  \\
&=& \Gamma_0^\mu(\{p_j^\prime\}, \{p_j\}) = \sum _j
\Gamma_{0,j}^\mu(\{p_j^\prime\}, \{p_j\})\nonumber   \\
&=& \eta _1\cdot \gamma^\mu\otimes S_F^{-1}(p_2)\otimes
S_F^{-1}(p_3)\otimes\dots\otimes S_F^{-1}(p_N) \cdot \hat\delta
_1({\{p_j^\prime-p_j\}}) \nonumber \\
&+& \eta _2\cdot S_F^{-1}(p_1)\otimes\gamma^\mu\otimes
S_F^{-1}(p_3)\otimes\dots\otimes S_F^{-1}(p_N) \cdot \hat\delta
_2({\{p_j^\prime-p_j\}}) \nonumber \\
&+& \eta _3\cdot
S_F^{-1}(p_1)\otimes S_F^{-1}(p_2)\otimes\gamma^\mu\otimes\dots\otimes
S_F^{-1}(p_N) \cdot \hat\delta 
_3({\{p_j^\prime-p_j\}}) \nonumber \\ &\vdots& \nonumber \\
&+& \eta _N\cdot S_F^{-1}(p_1)\otimes S_F^{-1}(p_2)\otimes
S_F^{-1}(p_3)\otimes\dots\otimes \gamma^\mu \cdot \hat\delta
_N({\{p_j^\prime-p_j\}}) \nonumber
\end{eqnarray}

with $i=1,2,\dots, N-1$ and $j,k=1,2,\dots, N$ and the short--hand
notation

\begin{eqnarray}
\hat\delta _k({\{p_j^\prime-p_j\}}) := (2\pi)^{4(N-1)} \cdot\delta^4(p_1^\prime-p_1)
\delta^4(p_2^\prime-p_2)\dots\delta^4(p_{k-1}^\prime-p_{k-1})
\delta^4(p_{k+1}^\prime-p_{k+1})\dots\delta^4(p_N^\prime-p_N) \nonumber  \quad .
\end{eqnarray}

The coordinate choice $\eta _j=e_j/{\cal Q}$ induces a re--definition
of the free vertex in eq. (\ref{GammaMuExplicitly}) by

\begin{eqnarray}
\label{VertexRedefinition}
\Gamma_0^\mu = \sum _{i=1}^N\Gamma_{0,i}^\mu \quad\longrightarrow\quad
\frac{1}{\cal Q}\Gamma_0^\mu = \frac{1}{\cal Q}\sum
_{i=1}^N\Gamma_{0,i}^\mu 
\end{eqnarray}

such that now 

\begin{eqnarray}
\Gamma_{0,i}^\mu = e _j \cdot S_F^{-1}(p_1)\otimes\dots\otimes
S_F^{-1}(p_{j-1})\otimes\gamma^\mu\otimes
S_F^{-1}(p_{j+1})\otimes\dots\otimes S_F^{-1}(p_N) \cdot \hat\delta 
_j({\{p_j^\prime-p_j\}})\nonumber 
\end{eqnarray}

holds. This procedure can also be applied to the full one--photon
vertex $\Gamma^\mu$ where also contributions from the one--photon
irreducible interaction kernel $K^\mu$ are included; in all terms that
are considered the photon
can be coupled to each fermion line in the corresponding
diagram. Analogous re--definitions for the two--photon vertices, {\it
i.e.} 

\begin{eqnarray}
\label{VertexRedefinitionmunu}
\Gamma^\munu  \quad\longrightarrow\quad\frac{1}{\cal Q}\Gamma^\munu 
\end{eqnarray}

can also be given. We
stress that this scheme is essentially equivalent to the introduction
of a proper electromagnetic charge operator $\textsf{Q}_f$ by hand
followed by an evaluation of the contraction of all indices in flavour space.

In the following subsections, we will give --- together with the general
results --- also the explicit lowest order formulae for $q\bar q$
mesons to make clear that these re--definitions give the correct and complete
matrix elements for electromagnetic form factors and Compton scattering.

\subsection{Current Normalization}
                                    \label{CurrentNorm} 

From Lorentz invariance and charge conjugation symmetry, one can
deduce the matrix element of the electromagnetic current to be
of the on--shell form $J^\mu={\cal Q}\cdot (P+P')^\mu f(q^2)$ where $f(q^2)$ is the
form factor with the normalization $f(0)=1$ and $\cal Q$ is the total
charge of the bound state. In the limit of vanishing
photon momentum $q=P'-P\to 0$, we therefore find 

\begin{eqnarray}
\lim_{q\to 0} J^\mu  = \lim_{q\to 0}\langle P' \left| j^\mu (0)
\right|P\rangle = {\cal Q}\cdot 2P^\mu \quad .\nonumber 
\end{eqnarray}

Let us now consider the normalization condition of the
Bethe--Salpeter amplitude describing a bound state of momentum $P$ and
mass $M$ (see {\it e.g.} \cite{Lurie}):

\begin{eqnarray}
\bar\chi (P) \Big( \frac{\partial}{\partial P_\mu}
\underbrace{\big(G_0^{-1}(P)+iK(P)\big)}_{=G^{-1}(P)} 
\left.\Big)\right|_{P^2=M^2} \chi (P) = i \: 2P^\mu\quad .\nonumber 
\end{eqnarray}

With the Ward--Takahashi in eq. (\ref{delGamma1stOrder}), recalling
$G^{-1}=:\Gamma$  and by
comparison with the definition of the current in  terms of Bethe--Salpeter amplitudes and
the one--photon vertex --- see eq. (\ref{CurrentDef}) ---, we easily
find the correct form factor normalization for $q\to 0'$:

\begin{eqnarray}
\label{BSEFormFactorNorm}
 -\bar\chi(P){ \Gamma}^\mu(P,P)\chi(P)={\cal Q}\cdot 2P^\mu \quad .
\end{eqnarray}

The total charge $\cal Q$ on the right--hand side comes from the
re--definition of the vertex in eq. (\ref{VertexRedefinition}) acting
such that a modified Ward--Takahashi identity $\partial^\mu
G^{-1}=-i\frac 1 {\cal Q} \Gamma^\mu$ has to be applied.

This low--energy limit is even correct if we neglect contributions
from the one--photon irreducible interaction kernel $K^\mu$. Due to
the Ward identity $\partial^\mu K=iK^\mu$, the
interaction kernel $K$ then must be independent of the total momentum
$P$ (as we have also found for the gauge invariance of the lowest
order current $J_0^\mu$). We thus can apply the Ward--Takahashi
identity in eq. (\ref{delFreeGamma1stOrder}) for the free one--photon vertex: 

\begin{eqnarray}
\label{BSEFormFactorNorm0}
 -\bar\chi(P){ \Gamma}_0^\mu(P,P)\chi(P)={\cal Q}\cdot 2P^\mu \quad .
\end{eqnarray}

As an example, we give the left--hand side of this last expression explicitly for a $q\bar q$
bound state and a non--vanishing photon momentum $q\not =0$ (note that
here the anti--quark is defined with  momentum $-p_2$ and  charge $-e_2$):

\begin{eqnarray}
\left.J^\mu\right|_{q\bar q} = &-e_1&\int\frac{d^4 p}{(2\pi)^4}\;
\bar\chi(P,p+\eta _2 q)\Big(\gamma^\mu\otimes S_F^{-1}(-\eta _2 P
+p)\Big)\chi(P,p)\nonumber  \\
&+e_2&\int\frac{d^4 p}{(2\pi)^4}\;
\bar\chi(P,p-\eta _1 q )\Big(S_F^{-1}(\eta _1 P
+p)\otimes\gamma^\mu\Big)\chi(P,p) \nonumber \quad .
\end{eqnarray}

The charge factors introduced in the preceeding subsection are
obviously correct; the indices of $\bar\chi{ \Gamma}^\mu_0\chi$
therefore must only be contracted in Dirac space (or, to be precise,
the contraction in flavour space will yield a factor $1$ due to the
unaltered charge operator $\textsf{Q}_f=\Id$).

\subsection{Born--Thomson Limit}
                                    \label{BornThomsonLimit} 

For charged particles, the Compton scattering tensor includes not only
structure dependent 
terms but also pole contributions in order ${\cal O}(\omega^0)$
with $\omega = (q_1, q_2)$; then
the Born part of $T^\munu$ in soft--photon approximation is known to
be (see {\it e.g.} \cite{SchererFearing}) 

\begin{eqnarray}
\label{BornThomsonTerm}
T^\munu_{\mbox{\footnotesize\sc Born}} &=&  {\cal Q}^2 \cdot 
f(q_1^2)f(q_2^2)\left[\frac{(2P+q_1)^\mu(2P'+q_2)^\nu}{s-M^2}
+ \frac{(2P'-q_1)^\mu(2P-q_2)^\nu}{u-M^2} - 2 g^{\mu\nu}\right] 
\end{eqnarray}

with the Mandelstam variables $s=(P+q_1)^2=(P'+q_2)^2$ and
$u=(P-q_2)^2=(P'-q_1)^2$ and the one--photon form factors $f(q_i^2)$
defined as in the preceeding subsection. Note that
$T^\munu_{\mbox{\footnotesize\sc Born}}$ is explicitly 
gauge invariant due to the non--singular term proportional to
$g^\munu$; the three terms in eq. (\ref{BornThomsonTerm}) can be
identified with the diagrams (a), (b) and (c) in fig. (\ref{fig:BornTerms}). 
The factor ${\cal Q}^2$ forces the amplitude to vanish for
neutral bound states in the low--energy limit.

\protect\begin{figure}[h]
  \protect\begin{center}
    \leavevmode
       \protect\input{BornTerms.pstex_t}
\newline
       \protect\caption{Born terms for Compton scattering off a spin--0 particle.}
       \protect\label{fig:BornTerms}
   \end{center}
\end{figure}

In the non--relativistic limit ({\it i.e.} for $q_i\to 0$), only the
Born terms in the full tensor $T^\munu$ survive. The amplitude 
${\cal T}$ for Compton scattering off a bound state of
mass $M$ with spacelike photon polarization vectors $\varepsilon
_i=(0,\vec\varepsilon _i)$ (and therefore $q_i\cdot\varepsilon _i=0$
and $P\cdot\varepsilon _i=P'\cdot\varepsilon _i=0$ for real photons)
then yields the celebrated result of Low, Gell--Mann and Goldberger
(see \cite{Thirring,Low,GellMannGoldberger}): 

\begin{eqnarray}
\label{BornThomsonAmplitude}
\lim_{q_i\to 0}{\cal T} = - \frac{1}{2M}\:\lim_{q_i\to
0}\:\varepsilon _{1\mu}T^\munu\varepsilon _{2\nu} =
-\frac{{\cal Q}^2}{M} \left(\vec\varepsilon _1\cdot\vec\varepsilon _2\right)\quad .
\end{eqnarray}

Obviously, the pole terms in eq. (\ref{BornThomsonTerm}) do not
contribute to the Compton amplitude in the very--low energy limit;
note furthermore that here and in the following we have used $f(0)=1$
for real photons.

Now we want to show that the full two--photon vertex $\Gamma^\munu$ in
fact yields this limit for $q_1,q_2\to 0$ ($P'\to P$). Let us
therefore consider eq. (\ref{BSEFormFactorNorm}) and make some remarks
on the right--hand side of this current normalization. Formally, the
factor $2P^\mu$ originates from a differentiation of the free boson
propagator $\Delta _F(P)=(P^2-M^2)^{-1}$ and a subsequent
amputation of this vertex in the limit $q\to 0$ (see
{\it e.g.} \cite{BjorkenDrell}):

\begin{eqnarray}
\tilde J^\mu :=\lim _{q\to 0} J^\mu = \Delta _F^{-1}(P) \left[\left(-{\cal Q}\frac{\partial}{\partial
P_\mu}\right) \Delta _F(P)\right]\Delta _F^{-1}(P) = {\cal Q}\cdot\frac{\partial}{\partial
P_\mu} \Delta _F^{-1}(P) = {\cal Q}\cdot 2P^\mu \quad .
\end{eqnarray}

The result is an amputated (point--form) boson--photon vertex; if we once more take
the derivative $-{\cal Q}\partial/\partial P_\nu$ then we end up with a two--photon vertex of
non--singular form:

\begin{eqnarray}
\tilde T^\munu_{\mbox{\footnotesize\sc
NoPole}} :=\lim _{q_i\to 0} T^\munu_{\mbox{\footnotesize\sc
NoPole}} = -{\cal Q}\cdot\frac{\partial}{\partial 
P_\nu}\tilde J^\mu =  -{\cal Q}^2\cdot\frac{\partial}{\partial
P_\nu}\frac{\partial}{\partial
P_\mu} \Delta _F^{-1}(P) = -{\cal Q}^2\cdot 2g^\munu \quad ;
\end{eqnarray}

see diagram (c) in fig. (\ref{fig:BornTerms}). $\tilde T^\munu_{\mbox{\footnotesize\sc
NoPole}}$ is equivalent to the non--singular part of
$T^\munu_{\mbox{\footnotesize\sc Born}}$ in the limit $q_i,q_i^2\to
0$, see eq. (\ref{BornThomsonTerm}). It can be recovered in the
Bethe--Salpeter formulation by taking the derivative of the
left--hand side of eq. (\ref{BSEFormFactorNorm}) and applying
$-{\cal Q}\partial/\partial P_\nu$ only to the amputated (point--form) two--photon vertex:

\begin{eqnarray}
\label{DerivativeChiGammamuChi}
\lefteqn{-\bar\chi(P)\left(-{\cal Q}\frac{\partial}{\partial
P_\nu}\Gamma^\mu(P,P)\right) \chi(P)} \\
&=& \lim _{q_i\to 0}
i\bar\chi(P')\Big(\Gamma^\munu(P',P)-\Gamma^\nu(P',P'+q_2)G(P'+q_2)
\Gamma^\mu(P+q_1,P)\nonumber\\&&
\phantom{ \lim _{q_i\to 0}
i\bar\chi(P')\Big(\Gamma^\munu(P',P)} - \Gamma^\mu(P',P'-q_1)G(P-q_2)\Gamma^\nu(P\:-q_2,P)\Big) \chi(P)
\nonumber\\&=&\lim _{q_i\to 0} \nonumber 
i\bar\chi(P')\Gamma_{\mbox{\footnotesize\sc NoPole}}^\munu(P',P)  \chi(P)
\end{eqnarray}

Here we have used that $\partial^\nu \Gamma ^\mu = \frac{i}{\cal Q}
\Gamma^\munu_{\mbox{\footnotesize\sc NoPole}}$, see
eq. (\ref{delGamma2ndOrder}) and eq. (\ref{VertexRedefinitionmunu}) for the
vertex' re--definition. In analogy to eq. (\ref{VertexSplit}), 
we now split the Compton scattering tensor into two parts by defining
$T^\munu=T^\munu_{\mbox{\footnotesize\sc Pole}} +
T^\munu_{\mbox{\footnotesize\sc NoPole}}$ where
$T^\munu_{\mbox{\footnotesize\sc Pole}}$ includes all pole
terms. Comparison with eq. (\ref{DefTmunu}) then yields

\begin{eqnarray}
\label{TmunuNoPoleLimit}
\lim _{q_i\to 0} T^\munu_{\mbox{\footnotesize\sc
NoPole}}  =\lim _{q_i\to 0}
i\bar\chi(P')\Gamma_{\mbox{\footnotesize\sc NoPole}}^\munu(P',P)
\chi(P)= -{\cal Q}^2\cdot 2g^\munu 
\end{eqnarray}

which is the correct low--energy limit for the non--singular terms in
the Compton scattering tensor. In the following, we will only work in
the limit $q_i\to 0$ ({\it i.e.} $P'\to P$), therefore skip the
dependence on the total four--momentum $P$ and use the abbreviation
$\partial^\nu:=\frac{\partial}{\partial P_\nu}$.
 
Let us now study the pole contributions in the low--energy limit
of $T^\munu$ which are necessary ingredients of the full tensor with
respect to gauge invariance. First, we will investigate the second
line of eq. (\ref{DerivativeChiGammamuChi}) and find that the terms
$\Gamma^\mu G \Gamma^\nu + \Gamma^\nu G \Gamma^\mu$ indeed reveal the
correct pole structure. This is due to the fact that the full Green's
function $G(\tilde P)$ has a pole for $\tilde P^2\to M^2$ (and
therefore in the limit of vanishing photon momenta) with the
residue $-i\chi\circ\bar\chi$, see eq. (\ref{GreensPole}). Recasting
the whole expression and using eq. (\ref{TmunuNoPoleLimit}) yields

\begin{eqnarray}
\lim _{q_i\to 0} \Big(
\bar\chi\Gamma^\mu\chi\cdot\frac{1}{s-M^2}\cdot\bar\chi\Gamma^\nu\chi
+\bar\chi\Gamma^\nu\chi\cdot\frac{1}{u-M^2}\cdot\bar\chi\Gamma^\mu\chi
+  T^\munu_{\mbox{\footnotesize\sc NoPole}} \Big) = \lim _{q_i\to 0}
i\bar\chi\Gamma^\munu\chi \qquad .
\end{eqnarray}

Since $J^\mu=-\bar\chi(P')\Gamma^\mu(P',P)\chi(P)={\cal Q}(P+P')^\mu
f(q^2)$ with $q=P'-P$ holds for the electromagnetic current, we
recover the full Born term $T^\munu_{\mbox{\footnotesize\sc Born}}$ of
eq. (\ref{BornThomsonTerm}) in the limit ${q_i\to 0}$ on
the left--hand side of this equation; it includes all pole terms plus a
non--singular contribution that restores gauge invariance. Only these
Born terms contribute to the low--energy limit of Compton scattering
because they are of zeroth order in the photon momenta; we therefore find

\begin{eqnarray}
\lim _{q_i\to 0} T^\munu = \lim _{q_i\to 0} \big( T^\munu_{\mbox{\footnotesize\sc
Pole}} +  T^\munu_{\mbox{\footnotesize\sc NoPole}} \big) = \lim
_{q_i\to 0} i\bar\chi\Gamma^\munu\chi 
\end{eqnarray}

and thus end up with the correct Born-Thomson limit for the Compton
scattering amplitude if we use the full two--photon irreducible vertex
$\Gamma^\munu$:  

\begin{eqnarray}
\lim_{q_i\to 0}\left.{\cal T}\right| _{\Gamma^\munu} = - \frac{1}{2M}\:\lim_{q_i\to
0}\:\varepsilon _{1\mu}T^\munu\varepsilon _{2\nu} =
-\frac{{\cal Q}^2}{M} \left(\vec\varepsilon _1\cdot\vec\varepsilon _2\right)\quad .\nonumber 
\end{eqnarray}

There is also an alternative way to demonstrate the correct
low--energy behaviour of the full two--photon irreducible vertex
$\Gamma^\munu$. We recall that we have only used results from
differentiations of amputated (point--form) operators such as $J^\mu$
and $\Gamma^\mu$. The pole terms can also be found if we define
$\tilde T^\munu$ in analogy to $\tilde
J^\mu$ as a twofold differentiation of the free
boson propagator $\Delta _F(P)$ with respect to the momentum $P$
followed by an amputation of the resulting vertex in the limit $q\to
0$ (see {\it e.g.} \cite{BjorkenDrell}):

\begin{eqnarray}
\tilde T^\munu :=\lim _{q_i\to 0} T^\munu = \Delta
_F^{-1} \left[\left(-{\cal Q}{\partial^\nu}\right)\left( -{\cal Q}{\partial
^\mu}\right)\Delta _F\right]\Delta _F^{-1} = \tilde
J^\mu \Delta _F\tilde J^\nu + J^\nu \Delta _F\tilde J^\mu - {\cal
Q}^2\cdot 2g^\munu
\end{eqnarray}

Inserting the explicit expressions for $J^\mu$ and $\Delta _F$, we
find the low--energy limit of the Born term in eq. (\ref{BornThomsonTerm}) including all
pole contributions. This procedure is equivalent to a differentiation of the full
expression $-\bar\chi\Gamma^\mu\chi$ in eq. (\ref{BSEFormFactorNorm});
in this way, we do not only take into account the derivative of the
amputated (point--form) vertex $\Gamma^\mu$ but also
include terms proportional to $\partial^\nu\chi$ and
$\partial^\nu\bar\chi$ which will then re--introduce the pole terms
that are also present in $\tilde T^\munu$. With the differential identities for $\partial
^\nu \Gamma^\mu$ and for the Bethe--Salpeter amplitudes in eqs. (\ref{delGamma2ndOrder}),
(\ref{delChi1stOrder}) and (\ref{delbarChi1stOrder}), and from the vertex' re--definition
in eq. (\ref{VertexRedefinitionmunu}), we recover

\begin{eqnarray}
{-\left(-{\cal Q}{\partial^\nu}\right)\bar\chi\Gamma^\mu\chi}
&=& {\cal Q}\big((\partial^\nu\bar\chi)\Gamma^\mu\chi
+\bar\chi(\partial^\nu\Gamma^\mu)\chi+\bar\chi\Gamma^\mu(\partial^\nu\chi)\big) \nonumber  \\
&=& i\big(\bar\chi\Gamma^\nu G\Gamma^\mu\chi
+\bar\chi(\Gamma^\munu-\Gamma^\mu G\Gamma^\nu - \Gamma^\nu G
\Gamma^\mu)\chi + \bar\chi\Gamma^\mu G \Gamma^\nu\chi\big) \nonumber
\\&=& i\bar\chi\Gamma^\munu\chi \nonumber\quad .
\end{eqnarray}

For the relevance of the formal differentiation of the
Bethe--Salpeter amplitudes $\chi$ and $\bar\chi$ which are properly
defined only on mass shell, see the remarks in section \ref{DiffWTI}. 
It is crucial for obtaining this correct result that the Ward identity of the order
${\cal O}(e^2)$ in eq. (\ref{delGamma2ndOrder}) includes subtraction terms that just
cancel the contributions from  $\partial^\nu\chi$ and
$\partial^\nu\bar\chi$. In other words:
the pole terms are re--introduced by the contributions coming from the
Bethe--Salpeter amplitudes in an external field, see fig. (\ref{fig:delChi1stOrder}).
In this sense we can state that a consistent differentiation prescription with regard to
point--form parts ({\it i.e.} $\tilde J^\mu$ and $\Gamma^\mu$) and
non--amputated parts (like Bethe--Salpeter amplitudes and boson
propagators that are cut {\it after} taking the derivatives) leads to
the correct Born--Thomson limit. Both ways yield, as shown above,

\begin{eqnarray}
\lim _{q_i\to 0} T^\munu = \lim _{q_i\to 0} \big( T^\munu_{\mbox{\footnotesize\sc
Pole}} +  T^\munu_{\mbox{\footnotesize\sc NoPole}} \big) = \lim
_{q_i\to 0} i\bar\chi\Gamma^\munu\chi \quad .
\end{eqnarray}

However, we shall note here that we prefer the first derivation which can be
formulated without using terms like $\partial^\nu\chi$ and
$\partial^\nu\bar\chi$ which are not strictly defined for on--shell
quantities like Bethe--Salpeter amplitudes; note that the relations in
eqs. (\ref{delChi1stOrder}) and
(\ref{delbarChi1stOrder}) have been found on quite formal grounds.

The Born--Thomson limit can also be derived if we only include
re--scattering terms due to $T$ matrix contributions but neglect
one--photon or two--photon irreducible interaction kernels. This can
be seen by applying the same procedure as before to the lowest order
current matrix element in eq. (\ref{BSEFormFactorNorm0}):

\begin{eqnarray}
\tilde T^\munu_{\mbox{\footnotesize\sc NoPole}} &=& -{\cal Q}^2\cdot
2g^\munu = -{\cal Q}\partial^\nu\tilde J^\mu = -\bar\chi\big(-{\cal
Q}\partial^\nu\Gamma^\mu_0\big)\chi \\ &=& i
\bar\chi(\Gamma_0^\munu-\Gamma_0^\mu G_0\Gamma_0^\nu - \Gamma_0^\nu G_0
\Gamma_0^\mu)\chi  \nonumber \\ &=& i
\bar\chi(\underbrace{\Gamma_0^\munu-i\Gamma_0^\mu T\Gamma_0^\nu - i\Gamma_0^\nu T
\Gamma_0^\mu}_{=\hat\Gamma^\munu}-\Gamma_0^\mu G\Gamma_0^\nu - \Gamma_0^\nu G
\Gamma_0^\mu)\chi \nonumber\qquad ;
\end{eqnarray}

here, we have only considered the differentiation of the amputated
(point--form) operators and used that $G=G_o-iT$. We thus find
$\hat\Gamma^\munu\to\frac{1}{\cal Q}\hat\Gamma^\munu$ as the 
modified tensor defined in eqs. (\ref{HatGreensFunction2Order2}),
(\ref{ModifiedGammamunu}) and (\ref{VertexRedefinitionmunu}) as well as pole
terms that can be re--arranged to the left--hand side. For $q_i \to
0$, the pole terms in the Green's function $G$ dominate all other
contributions; analogously to the derivation of the low--energy
limit of the full two--photon vertex, we are led to

\begin{eqnarray}
\lim _{q_i\to 0} \Big(
\bar\chi\Gamma_0^\mu\chi\cdot\frac{1}{s-M^2}\cdot\bar\chi\Gamma_0^\nu\chi
+\bar\chi\Gamma_0^\nu\chi\cdot\frac{1}{u-M^2}\cdot\bar\chi\Gamma_0^\mu\chi
+  T^\munu_{\mbox{\footnotesize\sc NoPole}} \Big) = \lim _{q_i\to 0}
i\bar\chi\hat\Gamma^\munu\chi \qquad .
\end{eqnarray}

Since the one--photon irreducible interaction current $K^\mu$ (which
is neglected in this approximation) does not
contribute in the static limit for interaction kernels $K$ that are
independent of the total momentum, see also
eq. (\ref{BSEFormFactorNorm0}), we find the correct low--energy
behaviour for the two--photon vertex $\hat\Gamma^\munu$:

\begin{eqnarray}
\lim _{q_i\to 0} T^\munu = \lim _{q_i\to 0} \big( T^\munu_{\mbox{\footnotesize\sc
Pole}} +  T^\munu_{\mbox{\footnotesize\sc NoPole}} \big) = \lim
_{q_i\to 0} i\bar\chi\hat\Gamma^\munu\chi \quad .
\end{eqnarray}

This leads to the Born--Thomson limit for the Compton amplitude
although we neglected contributions from $n$--photon irreducible
interaction kernels in $\hat T^\munu:=i\bar\chi\hat\Gamma^\munu\chi$: 

\begin{eqnarray}
\lim_{q_i\to 0}\left.{\cal T}\right| _{\hat\Gamma^\munu} = - \frac{1}{2M}\:\lim_{q_i\to
0}\:\varepsilon _{1\mu}\hat T^\munu\varepsilon _{2\nu} =
-\frac{{\cal Q}^2}{M} \left(\vec\varepsilon _1\cdot\vec\varepsilon _2\right)\quad .\nonumber 
\end{eqnarray}

Let us finally note that we cannot expect to recover the  
correct low--energy limit if we only consider the lowest order contribution to Compton
scattering in $T_0^\munu:=i\bar\chi\Gamma^\munu_0\chi$:

\begin{eqnarray}
\label{BornLimit0}
\lim_{q_i\to 0}\left.{\cal T}\right| _{\Gamma_0^\munu} = - \frac{1}{2M}\:\lim_{q_i\to
0}\:\varepsilon _{1\mu}T_0^\munu\varepsilon _{2\nu} \not=
-\frac{{\cal Q}^2}{M} \left(\vec\varepsilon _1\cdot\vec\varepsilon _2\right)\quad .\nonumber 
\end{eqnarray}

This is obvious from the considerations given above: if we do
not include pole contributions then we will not be able to produce the
full gauge invariant Born term, see
eq. (\ref{BornThomsonTerm}). However, in contrast to the $-{\cal
Q}\cdot 2g^\munu$ part, these pole terms do not
contribute to the Compton amplitude ${\cal T}$; but for the correct
reproduction of the limit $-\frac{{\cal Q}^2}{M}(\vec\varepsilon
_1\cdot\vec\varepsilon _2)$ we must at least include the full
non--singular tensor $T^\munu_{\mbox{\footnotesize\sc NoPole}}$ and
not only its lowest--order part, see eq. (\ref{TmunuNoPoleLimit}).

As an example, we will finally write down explicitly this
lowest--order Compton scattering tensor
$T_0^\munu=i\bar\chi\Gamma^\munu_0\chi$ for a $q\bar q$ system with
finite photon momenta $q_1,q_2\not =0$:

\begin{eqnarray}
\left.T^\munu _0 \right|_{q\bar q}=& \:\:\:-ie_1 e_2&\int\frac{d^4 p}{(2\pi)^4}\;
\bar\chi(P,p+\eta _2 q_1 + \eta _1 q_2) \Big( \gamma^\mu\otimes\gamma^\nu  \Big)\chi(P,p)\nonumber  \\
& -ie_1 e_2&\int\frac{d^4 p}{(2\pi)^4}\; \bar\chi(P,p-\eta _1 q_1 - \eta _2 q_2) \Big(
\gamma^\nu\otimes\gamma^\mu  \Big)\chi(P,p) \nonumber \\ 
& +ie_1^2&\int\frac{d^4 p}{(2\pi)^4} \;\bar\chi(P,p+\eta _2 q_1 - \eta _2 q_2) \Big(
 \gamma^\nu S_F(\eta _1 P +p+q_1)\gamma^\mu\otimes S_F^{-1}(-\eta _2 P +p)
\Big)\chi(P,p) \nonumber \\ 
& +ie_1^2&\int\frac{d^4 p}{(2\pi)^4}\; \bar\chi(P,p+\eta _2 q_1 - \eta _2 q_2) \Big(
 \gamma^\mu S_F(\eta _1 P +p-q_2)\gamma^\nu\otimes S_F^{-1}(-\eta _2 P +p)
\Big)\chi(P,p) \nonumber \\ 
& +ie_2^2&\int\frac{d^4 p}{(2\pi)^4} \;\bar\chi(P,p-\eta _1 q_1 + \eta _1 q_2) \Big(
S_F^{-1}(\eta _1 P +p)\otimes\gamma^\nu S_F(-\eta _2 P+p-q_1)\gamma^\mu 
\Big)\chi(P,p) \nonumber \\ 
& +ie_2^2&\int\frac{d^4 p}{(2\pi)^4}\; \bar\chi(P,p-\eta _1 q_1 + \eta _1 q_2) \Big(
S_F^{-1}(\eta _1 P +p)\otimes\gamma^\mu S_F(-\eta _2 P+p+q_2)\gamma^\nu 
\Big)\chi(P,p) \nonumber  \quad .
\end{eqnarray}

Again, the correct charge factors are found for the different terms contributing
to $T^\munu _0$; note that the anti--quark with momentum $-p_2=-P/2+p$
has the charge $-e_2$. Six diagrams emerge from this lowest order two--photon vertex;
if $T$ matrix contributions are taken into account and interaction
currents are still neglected, eight additional terms will
have to be considered for a $q\bar q$ system.

\section{Summary and Conclusions}
                                    \label{Summary} 

We have derived the second order Green's function $G^\gammagamma$ for Compton
scattering off a bound state in a scheme that the authors of
ref. \cite{Blankleider1,Blankleider2,Blankleider3} (where it has been introduced for processes in 
first order of the electromagnetic coupling) have labelled ``gauging a hadronic
system''. The resulting full two--photon vertex $\Gamma^\gammagamma =
G^{-1}G^\gammagamma G^{-1}$ includes the lowest order contribution
$\Gamma^\gammagamma_0$ and an explicit two--photon irreducible
interaction kernel $K^\gammagamma$ but also re--scattering terms via
an intermediate full Green's function $\Gamma ^\gamma G \Gamma
^\gamma$ (with $G=G_0-iT$) as well as subtraction terms $\Gamma _0^\gamma G_0 \Gamma
^\gamma_0$. Note that the singularities in $\Gamma ^\gamma G \Gamma
^\gamma$ due to the poles in the Green's function for vanishing photon
momenta are responsible for the correct low--energy limit of the
Compton scattering tensor including the pole contributions present in
the Born terms.

By imposing that the amplitude of order ${\cal O}(e^2)$ obeys the
related Ward--Takahasi identity, we found a constraint 
for the two--photon irreducible interaction kernel similar to the one
in first order (see also \cite{GrossIto}). We also dicussed the
Ward--Takahashi identities for the ``gauged'' Bethe--Salpeter
amplitudes $\chi^\mu$ and $\chi^\munu$. Differential identities for
Green's functions, the vertices $\Gamma^\mu$ and $\Gamma^\munu$ and
the $n$--photon irreducible interaction kernels were derived. By using
rather formal arguments, we found similar identities for the
Bethe--Salpeter amplitudes; however, we stressed that these are
strictly defined only on mass shell, {\it i.e.} for $P^2=M^2$.

As we checked the gauge invariance condition for a matrix element
proportional to $\bar\chi\Gamma^\gammagamma \chi$, we found that it is
(as it should be) automatically satisfied if the two--photon vertex
obeys the correct Ward--Takahashi identity. For an approximation in
which we neglected all $n$--photon irreducible interaction kernels, we
found that gauge invariance for the resulting two--photon vertex
$\hat\Gamma^\gammagamma$ is preserved only if the Bethe--Salpeter
kernel $K$ does not depend on the bound state's four--momentum $P$ and
is of local type, {\it i.e.} $K(P;\{k_i\}, \{k_i'\})= K(\{k_i -
k_i'\})$ where $\{k_i\}$ denote the relative momenta of the
constituents. This is the same condition as for the related
approximation in first order if we neglect the one--photon irreducible
kernel, {\it i.e.} $\Gamma^\gamma =\hat\Gamma^\gamma =\Gamma^\gamma_0$. 

A similar result could be derived with regard to the low--energy
limits: they are properly described by the full vertices
$\Gamma^\gamma$ and $\Gamma^\gammagamma$. If we only neglect contributions
from $K^\gamma$ and $K^\gammagamma$, then we also meet the correct
limits for vanishing photon momenta. However, we do not find the
correct Born--Thomson limit for the Compton scattering amplitude if we
only include the lowest order irreducible vertex
$\Gamma^\gammagamma_0$. Therefore we conclude that the corresponding
vertex to the lowest order vertex $\hat\Gamma^\gamma=\Gamma^\gamma_0$ at ${\cal O}(e)$
is in second order of the electromagnetic coupling not
$\Gamma^\gammagamma_0$ but merely
$\hat\Gamma^\gammagamma=\Gamma^\gammagamma_0 - i \Gamma_0^\gamma T
\Gamma^\gamma_0$ as a vertex that also includes re--scattering terms due
to $T$ matrix contributions. 

Let us finally note that the scheme presented in this article can also
be applied to processes like $\gammagamma\to {\cal A}\bar{\cal A}$,
{\it i.e.} the production of bound state pairs ({\it e.g.} $\pi^+\pi^-$ or
$\pi^0\pi^0$) in photon--photon collisions. The corresponding
identities could as well be obtained by introducing new momentum
variables $q_2\to -q_2$ and $P\to -P$ that satisfy the four--momentum conservation
$q_1+q_2=P+P'$. Furthermore, we want to stress that our results
could also be of interest for Compton scattering off an
electromagnetic bound state such as positronium since we didn't specify
the interaction kernel in the Bethe--Salpeter equation, and in this
respect our considerations are quite general.

\vspace{1cm}


\section*{Acknowledgements}

We thank H. R. Petry, B. Ch. Metsch and K. Kretzschmar for
stimulating and fruitful discussions. Financial support of the {\sc Deutsche
Forschungsgemeinschaft} (DFG) is gratefully acknowledged.


\vspace{1cm}

\bibliographystyle{plain}

\vspace{2cm}


\begin{appendix}

\section{Coordinates for a Composite System}        	   
                                        \label{app:Coordinates} 
			
Let us consider a system of $n$ fermions and $\bar n$ anti--fermions
($N=n+\bar n$) where the position of each constituent is described by
the coordinate $x_i$ ($i=1,2,\dots, N$). Now we introduce new
coordinates by

\begin{eqnarray}
\label{XTrafo}
X &:=& \eta _1 x_1 + \eta _2 x_2 + \dots + \eta _N x_N \\
r_1 &:=& x_1 - x_2 \nonumber \\
r_2 &:=& \frac 1 2 (x_1 + x_2) - x_3 \nonumber \\
&\vdots&\nonumber \\
r_{N-1} &:=& \frac{1}{N-1}(x_1 + x_2 +  + \dots + x_{N-1} ) - x_N \nonumber\quad .
\end{eqnarray}

The coefficients $\eta _i$ have no direct geometrical meaning; the
only constraint is that they have to sum up to unity:

\begin{eqnarray}
\eta _1 + \eta _2 + \dots + \eta _N = 1 \quad .\nonumber 
\end{eqnarray}

Choosing them as $\eta _i=m_i/M$ with $M=\sum _i m_i$ gives $X$ the meaning of a
center--of--mass coordinate; however, a more common choice is simply $\eta
_i=1/N$. Another possibility is to interpret $X$ as a
center--of--charge coordinate by adopting $\eta _i=e_i/{\cal Q}$
where $e_i$ is the physical charge of the $i$--th constituent and
${\cal Q}=\sum _i e_i$ is the total charge; this choice will be
applied in the last section of this article. Note that the fixing
$\eta _i=e_i/{\cal Q}$ is formally also possible for neutral bound
states since even for ${\cal Q}=0$ the relation $\sum _i \eta_i=1$ is satisfied.

In a compact form, the transformation in eq. (\ref{XTrafo}) can be written as a matrix
equation like

\begin{eqnarray}
\label{XTrafoMatrix}
\alpha _x = \textsf{A}_x \beta _x \qquad &\mbox{with}& \quad
\alpha _x = \left(
\begin{array}{c}
X \\ r_1 \\ \vdots \\ r_{N-1}
\end{array}
\right)\quad , \quad 
\beta _x = \left(
\begin{array}{c}
x_1 \\ x_2 \\ \vdots \\ x_N
\end{array}\right) \\
 &\mbox{and}& \quad\textsf{A}_x = \left(
\begin{array}{ccccc}
\eta _1 &\eta _2 &\eta _3 &\dots &\eta _N \\
1 & -1 & 0 &\dots & 0 \\
\frac 1 2  & \frac 1 2 & -1 &\dots & 0 \\
\vdots & \vdots & \vdots &\ddots & \vdots \\
\frac{1}{N-1} & \frac{1}{N-1} & \frac{1}{N-1}&\dots & -1 \\
\end{array}\right)\quad . \nonumber 
\end{eqnarray}

Now consider the canonical conjugated momenta $p_1, p_2, \dots, p_N$ and
$P,k_1, \dots, k_{N-1}$, respectively, where $P$ is the total momentum
$P=p_1+p_2+\dots + p_N$. We can give a matrix equation for this
transformation in momentum space as well:

\begin{eqnarray}
\label{PTrafoMatrix}
\alpha _p = \textsf{A}_p \beta _p \qquad &\mbox{with}& \quad
\alpha _p = \left(
\begin{array}{c}
P \\ k_1 \\ \vdots \\ k_{N-1}
\end{array}
\right)\quad , \quad 
\beta _p = \left(
\begin{array}{c}
p_1 \\ p_2 \\ \vdots \\ p_N
\end{array}\right)\quad .
\end{eqnarray}

Obviously, the $i$--th element in the first column of
$\textsf{A}_p^{-1}$ gives the dependency of the momentum $p_i$ of the
total momentum $P$.

The transformations given in eqs. (\ref{XTrafoMatrix}) and
(\ref{PTrafoMatrix}) satisfy the following relations
concerning their Jacobi determinant and their scalar product:

\begin{eqnarray}
\left| \frac{\partial(X, r_1,\dots,
r_{N-1})}{\partial(x_1,x_2,\dots,x_N)}\right| = \left|
\frac{\partial(P, k_1,\dots,
k_{N-1})}{\partial(p_1,p_2,\dots,p_N)}\right| = 1 \nonumber 
\end{eqnarray}

and

\begin{eqnarray}
PX + r_1 k_1 + \dots r_{N-1} k_{N-1} = p_1 x_1 + p_2 x_2 + \dots + p_N x_N\quad .
\end{eqnarray}

The last equation gives the transformation matrix
$\textsf{A}_p$ in momentum space in terms of the matrix
$\textsf{A}_x$ in position space:

\begin{eqnarray}
\alpha^t _p \alpha _x = \beta^t_p \beta _x \qquad\Longleftrightarrow\qquad
\alpha^t _p\textsf{A}_x \beta _x = (\textsf{A}_p^{-1}\alpha _p)^t \beta _x = \alpha^t _p
(\textsf{A}_p^{-1})^t \beta _x \quad .\nonumber 
\end{eqnarray}

This yields $\textsf{A}_p^{-1}=\textsf{A}_x^t$ so that we find the
dependency of the momentum of the $i$--th constituent of the total
momentum to be

\begin{eqnarray}
\label{pPRelation}
p_i = \eta _i P + f(k_1, \dots , k_{N-1})
\end{eqnarray}

where $f(k_1, \dots , k_{N-1})$ is a linear function that can be
derived from $\textsf{A}_p^{-1}=\textsf{A}_x^t$. The constituents of a
$q\bar q$ meson for instance have the momenta 
  
\begin{eqnarray}
\label{p1p2_def}
p_1 = \eta _1 P + p \qquad\mbox{and}\qquad -p_2 = -\eta _2 P + p 
\end{eqnarray}

with total momentum $P=p_1+p_2$ and relative momentum $p:=k_1=\eta _2 p_1 -
\eta _1 p_2$.

\section{Bethe--Salpeter Amplitudes in Second Order}        	   
                                        \label{app:BSA2}		   
								   
In this appendix, we want to derive the Bethe--Salpeter amplitude in
second order of an external field. We start with
eq. (\ref{BSExpansion}) and write down the expansion in
eq. (\ref{Expansion}) for $\textsf{O}=G_0, K, \chi$:

\begin{eqnarray}
\chi -ie\chi^\gamma - e^2\chi ^\gammagamma + \dots = -i 
(G_0 -ie G_0^\gamma - e^2 G_0 ^\gammagamma + \dots)
(K -ieK^\gamma - e^2K ^\gammagamma + \dots)
(\chi -ie\chi^\gamma - e^2\chi ^\gammagamma + \dots)\quad .\nonumber 
\end{eqnarray}

In the order ${\cal O}(e^2)$ of the electromagnetic coupling, we find
with the first order relation $\chi^\gamma = G\Gamma^\gamma\chi$

\begin{eqnarray}
\chi ^\gammagamma  &=& -i G_0 ^\gammagamma K\chi-i G_0
K^\gammagamma\chi-i G_0 K\chi^\gammagamma -iG_0^\gamma K^\gamma
\chi-iG_0^\gamma K \chi^\gamma-iG_0 K^\gamma 
\chi^\gamma \nonumber \\
 &=& \underbrace{(\Id+iG_0K)^{-1}}_{=GG_0^{-1}} \: \Big(-iG_0
\Gamma^\gammagamma \underbrace{G_0 K\chi}_{=i\chi}-i G_0
K^\gammagamma\chi-iG_0\Gamma_0^\gamma G_0 K^\gamma
\chi-iG_0\Gamma_0^\gamma \underbrace{G_0 K
G}_{=i(G-G_0)}\Gamma^\gamma \chi-iG_0 K^\gamma G\Gamma^\gamma \chi\nonumber\Big) \\
 &=& G\Big(\Gamma_0^\gammagamma - iK^\gammagamma - \Gamma _0^\gamma
G_0(\Gamma^\gamma+iK^\gamma) + (\Gamma _0^\gamma - iK^\gamma)G\Gamma^\gamma\Big)\chi\nonumber\\
 &=& G\Big(\Gamma_0^\gammagamma - iK^\gammagamma +
\Gamma^\gamma G\Gamma^\gamma - \Gamma _0^\gamma
G_0 \Gamma _0^\gamma\Big)\chi\nonumber 
\end{eqnarray}

where we have used the fundamental equation for the Green's function,
the Bethe--Salpeter equation and some relations introduced in
section \ref{SecondOrder}. With $\Gamma^\gammagamma :=
\Gamma_0^\gammagamma -i K^\gammagamma + \Gamma^\gamma G \Gamma^\gamma
- \Gamma_0^\gamma G_0 \Gamma_0^\gamma $, we finally find the expression in
eq. (\ref{BSA2ndOrder}) for the second order Bethe--Salpeter amplitude $\chi
^\gammagamma = G\Gamma^\gammagamma \chi$.

\end{appendix}

\end{document}